\documentclass[preprint,review,12pt]{elsarticle}

\usepackage{amssymb}
\usepackage{amsmath,bm}
\usepackage{colortbl,booktabs}
\usepackage{tabularx}
\usepackage{threeparttable}
\usepackage{multicol,multirow}
\usepackage{subfigure}
\usepackage [font=footnotesize,labelfont=bf]{caption}
\usepackage{rotating}
\usepackage{tikz}
\usepackage{hyperref}

\usepackage{color}
\definecolor{R}{rgb}{1, 0, 0}
\definecolor{G}{rgb}{0, 0.5, 0}
\definecolor{B}{rgb}{0, 0, 1}

\definecolor{yjzhuCol}{rgb}{0.9, 0, 0.2}
\definecolor{czhangCol}{rgb}{0.9, 0, 0.2}
\definecolor{xyhuCol}{rgb}{0.9, 0, 0.9}

\journal{Elsevier}

\begin{document}

\begin{frontmatter}

\title{A splitting random-choice dynamic relaxation method for smoothed particle hydrodynamics}

\author{Yujie Zhu \fnref{eqnote}}
\ead{yujie.zhu@tum.de}
\author{Chi Zhang \fnref{eqnote}}
\ead{c.zhang@tum.de}
\author{Xiangyu Hu \corref{mycorrespondingauthor}}
\cortext[mycorrespondingauthor]{Corresponding author.}
\ead{xiangyu.hu@tum.de}
\address{Department of Mechanical Engineering, Technical University of Munich\\
85748 Garching, Germany}
\fntext[eqnote]{Yujie Zhu and Chi Zhang contributed equally to this work}

\begin{abstract}
For conventional smoothed particle hydrodynamics (SPH), 
obtaining the static solution of a problem is time-consuming.
To address this drawback, we propose an efficient dynamic relaxation method by
adding large artificial-viscosity-based damping into the momentum conservation equation.
Then, operator splitting methods are introduced to discretize 
the added viscous term for relaxing the time-step limit.
To further improve the convergence rate, a random-choice strategy is adopted, 
in which the viscous term is imposed randomly rather than at every time step.
In addition, to avoid the thread-conflict induced by applying shared-memory parallelization 
to accelerate implicit method,
a splitting cell-linked list scheme is devised.
A number of benchmark tests suggest that the present method helps systems 
achieve equilibrium state efficiently.
\end{abstract}


\end{frontmatter}
%
%
\section{Introduction}
\label{sec1}
For the numerical simulation of solid dynamic applications undergoing large strain,  
as an alternative approach to the conventional finite element method (FEM), 
the meshless methods have attracted more and more interests in recent years \cite{belytschko1996meshless,liu2005introduction, liu2010smoothed}.
One notable Lagrangian meshless method is smoothed particle hydrodynamics (SPH),
which was first introduced by Lucy \cite{lucy1977numerical} and Gingold and Monaghan \cite {gingold1977smoothed} for astrophysics applications.
In SPH method, discretization operators are approximated 
by using the particles within a neighborhood 
with the help of a weight or smoothing kernel function.
This distinct non-local feature makes SPH method can handle 
large deformations straightforwardly.
After its inception, SPH method has been extensively studied and applied 
to a broad range of problems including fluid dynamics \cite{monaghan1994simulating,hu2006multi,shao2006simulation,zhang2019weakly}, 
solid dynamics \cite{benz1995simulations,randles1996smoothed,monaghan2000sph}, 
fluid-structure interactions (FSI) \cite{antoci2007,zhang2020multi} 
and multi-physics applications \cite{zhang2020integrative}. 
In addition,
to further improve the numerical accuracy and address 
the problems of tensile instability and inconsistency 
\cite{swegle1995smoothed,morris1996analysis},
many variants of SPH method have been developed.
These modifications include 
the symmetric and normalization formulations \cite{monaghan1992smoothed,johnson1996artificial,randles1996smoothed},
the corrective smoothed particle method (CSPH) \cite{chen1999corrective,chen2000generalized,bonet2002simplified},
the finite particle method (FPM) \cite{liu2005modeling,liu2006restoring}
and the stress point algorithm \cite{dyka1995approach,dyka1997stress,randles2000normalized}.
More applications and their corresponding algorithms can be found 
in comprehensive reviews as Refs. \cite{liu2010smoothed,monaghan2012smoothed,ye2019smoothed}.

Although the accuracy and stability are achieved by SPH methods 
for the simulation of solid dynamics problems,
one has to note that obtaining static solution is very time-consuming
because SPH is a typical method with explicit time-stepping.
Therefore, it is necessary to find an efficient method to approximate 
the converged static or quasi-static displacement.
Besides resorting to traditional implicit methods,
which solve the entire system with iterative solvers,
it is attractive to apply dynamic relaxation, 
which was proposed for FEM method originally 
\cite{otter1966dynamic,belytschko2013nonlinear,jung2018dynamic},
because its explicit iterative algorithm is more efficient 
for large problems, especially in highly nonlinear cases.

Generally, 
the dynamic relaxation technique can be classified into two groups, 
i.e. viscous dynamic relaxation (VDR) and kinetic dynamic relaxation (KDR) \cite{rodriguez2011numerical,alamatian2012new,rezaiee2014fictitious}, 
according to the manner of applied damping. 
In VDR, an artificial viscous damping term is added into 
the equation of motion to accelerate the convergence.
One difficulty of VDR is that the solutions for nonlinear problems may be path dependent 
and vary with different damping ratio as noted in Ref. \cite{lin2014efficient}.
Also, the relaxation is slow if insufficient damping ratio is applied, 
on the other hand, too much damping would also hinder the system 
from achieving final steady state because the damped velocity can be very small.
In KDR, damping is imposed by resetting the velocity to zero at every time step 
or when kinetic energy reaches maximum.
Although 
the procedure is simple and no damping factor is required,
the momentum conservation property is not satisfied in the process of KDR
and the convergent rate is not optimal \cite{alamatian2012new,zardi2020new}.

The main objective of this paper is to propose a dynamic relaxation method, 
where efficient damping is introduced to SPH for obtaining the static solution.
Following VDR, an artificial viscous term is added into the momentum conservation equation 
to dissipate the velocity gradient of the system. 
To address the issue of computational inefficiency 
when applying explicit scheme to integrate the viscous term, 
we propose two operator splitting schemes, 
one is inspired by a method for highly dissipative system \cite{litvinov2010splitting}.
Since such methods treat the viscous term implicitly,   
they can be unconditional stable and relax the viscous time-step limit,
and allow very large damping ratio.
To provide a remedy to the issues of path dependency 
and large damping ratio encountered by the original VDR and KDR,
and further accelerate the convergence, 
we propose a random-choice strategy by which 
the artificial damping is only carried out at a small fraction of time steps randomly. 

For large scale SPH simulations, 
two parallelization strategies, 
viz. shared and distributed memories, 
are implemented for accelerating computational performance. 
For shared memory parallelization strategy, 
thread-conflict occurs when implicit scheme is applied. 
To address this issue, 
a thread-safe splitting cell-linked list (CLL) scheme is proposed for implementing shared memory parallelization to accelerate implicit scheme.
In our code, 
the open-source Threading Building Blocks (TBB) library \cite{contreras2008characterizing} developed by INTEL is applied for providing the share-memory parallel programming environment. 

The remainder of this paper is organized as follows.
Section 2 briefly reviews the governing equations and 
total Lagrangian SPH (TLSPH) formulations for solid dynamics.
The details of the present artificial damping method and 
the splitting CLL scheme are described in Section 3.
Numerical examples are presented and discussed in Section 4 and 
then concluding remarks are given in Section 5.
The corresponding code of this work are available on GitHub at \url{https://github.com/Xiangyu-Hu/SPHinXsys} \cite{zhang2020sphinxsys}.
%
%
\section{Preliminary}
\label{preliminary}
\subsection{Total Lagrangian solid dynamics}\label{subsec:governing equations}
In total Lagrangian formalism,
the conservation of mass and momentum for solid dynamics can be expressed as
\begin{equation}
	\begin{cases}
		\rho =\rho^0 J^{-1}	\\
		\rho^0 \frac{\text{d} \mathbf v}{\text{d} t} = \nabla^0 \cdot \mathbb P^T + \rho^0 \mathbf g	 \\
	\end{cases},
	\label{governing-equations}
\end{equation}
where $\rho$ is the density, Jacobian $J = \text{det} (\mathbb F)$, $\mathbf v$ the velocity, $\mathbf g$ body-force and $\mathbb P$ the first Piola-Kirchhoff stress tensor.
$\mathbb F$ is the deformation gradient and is defined as
\begin{equation}
	\mathbb F = \nabla^0 \mathbf u + \mathbb I,
	\label{deformation-gradient}
\end{equation}
where $\mathbb I$ denotes the identity matrix and $\mathbf u$ the displacement vector.
Note that the superscript $\left( \cdot \right)^0 $ denotes the quantities in the reference configuration thereafter.
For ideal elastic or Kirchhoff material, the first Piola-Kirchhoff stress tensor $\mathbb P$, 
can be evaluated by 
\begin{equation}
\mathbb P = \mathbb F \mathbb S.
\label{piola}
\end{equation}
Here, $\mathbb S$ denotes the second Piola-Kirchhoff stress tensor
and it can be evaluated via constitutive equation.
In particular, for isotropic and linear elastic materials, the constitutive equation can be simplified to
\begin{equation}
\begin{split}
\mathbb S &= K \text{tr}\left( \mathbb E\right) \mathbb I + 2G\left( \mathbb E-\frac{1}{3} \text{tr}\left( \mathbb E\right) \mathbb I \right) \\
&= \lambda \text{tr} \left( \mathbb E\right) \mathbb I + 2\mu \mathbb E,
\end{split}
\label{constitutive-equation}
\end{equation}
where $\lambda$ and $\mu$ are Lam$\acute{e}$ coefficients and
$\mathbb E$ represents the Green-Lagrangian strain tensor defined by
\begin{equation}
\mathbb E = \frac{1}{2} \left( \mathbb F^T \mathbb F - \mathbb I \right).
\label{strain-tensor-e}
\end{equation}
Also, $K = \lambda + 2\mu/3$ and $G = \mu$ denote the bulk and the shear modulus, respectively. 
They have the following relationship
\begin{equation}
 E = 2 G  \left( 1 + 2 \nu \right) = 3 K  \left( 1 - 2 \nu \right),
\label{modulus-relation}
\end{equation}
with $E$ representing the Young's modulus and $\nu$ the Poisson's ratio.
In this work, 
the neo-Hookean material with the stain energy density function
\begin{equation}
{\mathcal W} = \mu \text{tr} \left( \mathbb E \right) - \mu \text{ln} J + \frac{\lambda}{2} \left(\text{ln} J \right) ^2
\label{strain-density}
\end{equation}
is also considered to predict the nonlinear stress-strain behavior of materials undergoing large deformations.
The corresponding second Piola-Kirchhoff stress $\mathbb S$ can be evaluated by
\begin{equation}
\mathbb S = \frac{\partial {\mathcal W}}{\partial \mathbb E}.
\label{PK-stress}
\end{equation}
\subsection{Total Lagrangian SPH method}\label{subsec:tl-sph}
In TLSPH method, 
the kernel correction matrix $\mathbb B^0$ is defined as \cite{vignjevic2006sph}
\begin{equation}
\mathbb B^0_i = \left(- \sum_{j}^{N} V^0_j \mathbf r^0_{ij} \otimes \nabla^0_i W_{ij} \right) ^{-1},
\label{correction-matrix}
\end{equation}
where $ \mathbf r^0_{ij} = \mathbf r^0_i - \mathbf r^0_j $ and 
\begin{equation}
\nabla^0_i W_{ij} = \left( \frac{\partial W_{ij}}{\partial r_{ij}}\right) ^0 \mathbf e^0_{ij} 
\label{correction-matrix1}
\end{equation}
represents the gradient of the kernel function calculated at the initial reference configuration and $V^0_j$ denotes the particle volume.
Here, $\mathbf e^0_{ij} = \frac{\mathbf r^0_{ij}}{\left| \mathbf r^0_{ij}\right|}$. 
According to the equation of mass conservation Eq. \eqref{governing-equations},
the density can be evaluated directly as
\begin{equation}
\rho_i = \rho^0_i J^{-1}.
\label{evaluate-mass}
\end{equation}
Then, the equation of momentum conservation in Eq. \eqref{governing-equations} is discretized as
\begin{equation}
\frac{\text{d} \mathbf v_i}{\text{d} t}= \frac{2}{m_i}\sum_{j}^{N} V^0_i V^0_j  \tilde{\mathbb P}_{ij} \nabla^0_i W_{ij} + \mathbf g,
\label{evaluate-momentum}
\end{equation}
where
\begin{equation}
\tilde{\mathbb P}_{ij}=\frac{1}{2}\left( \mathbb P_i \mathbb B^0_i + \mathbb P_j \mathbb B^0_j\right) 
\label{averaged-Piola-Kirchhoff}
\end{equation}
denotes the inter-particle averaged first Piola-Kirchhoff stress and $m_i$ is the particle mass.
Here, 
the first Piola-Kirchhoff stress tensor is computed with the constitutive law where the deformation tensor $\mathbb{F}$ is updated by
\begin{equation}
	\frac{ d \mathbb{F}_i}{dt} = -\left( \sum_j V^0_j \mathbf{v}_{ij} \otimes \nabla^0_i W_{ij}  \right) \mathbb{B}^0_i,
\end{equation}
with $ \mathbf v_{ij} = \mathbf v_i - \mathbf v_j $ the relative velocity between particle $i$ and $j$.
\subsection{Time integration}\label{subsec:time integration}
Following the work of Zhang et al. \cite{zhang2020multi},
the position-based Verlet scheme is adopted for time integration.
First, update the deformation tensor, density and particles position to a midpoint stage as 
\begin{equation}
\begin{cases}
\mathbb F^{n+\frac{1}{2}}_i = \mathbb F^n_i + \frac{1}{2} \Delta t \left( \frac{\text d \mathbb F_i}{\text d t} \right)^n\\
\rho^{n+\frac{1}{2}}_i =  \rho^{0}_i \left( J^n_i \right) ^{-1}  \\
\mathbf r^{n+\frac{1}{2}}_i = \mathbf r^{n}_i + \frac{1}{2} \Delta t \mathbf v_i^{n} \\
\end{cases}.
\label{timestep-1}
\end{equation}
Then, particles velocity at new time step is update by
\begin{equation}
\mathbf v^{n+1}_i =  \mathbf v^{n}_i +\Delta t \left( \frac{\text{d} \mathbf v_i}{\text{d}t} \right) ^ {n}.
\end{equation}
Finally, the deformation tensor, density as well as positions of particles at the new stage are evaluated by
\begin{equation}
\begin{cases}
\mathbb F^{n+1}_i = \mathbb F^{n+\frac{1}{2}}_i + \frac{1}{2} \Delta t \left( \frac{\text d \mathbb F_i}{\text d t} \right)^{n+\frac{1}{2}}\\
\rho^{n+1}_i =  \rho^{0}_i \left( J^{n+\frac{1}{2}}_i \right) ^{-1}  \\
\mathbf r^{n+1}_i = \mathbf r^{n+\frac{1}{2}}_i + \frac{1}{2} \Delta t \mathbf v_i^{n+\frac{1}{2}} \\
\end{cases}.
\label{timestep-3}
\end{equation}
For numerical stability, the acoustic and body-force time-step criterion are considered, i.e.
\begin{equation}
\Delta t = 0.6 \text{min} \left( \frac{h}{c+\left| \mathbf v \right|_{max} }, \sqrt{\frac{h}{\left|\frac{\text d \mathbf v}{\text d t} \right|_{max}}} \right),
\label{time-step-criterion}
\end{equation}
where $h$ denotes the smoothing length and $c$ is the sound speed of solid structure given by $c=\sqrt{K/\rho}$.
%
%
\section{Dynamic relaxation method}\label{dynamic relaxation}
To damp the velocity gradient of a system, 
an artificial viscous force can be introduced into the momentum conservation equation Eq. \eqref{governing-equations}.
Subsequently, 
the momentum conservation equation for solid dynamics can be rewritten as 
\begin{equation}
\rho^0 \frac{\text{d} \mathbf v}{\text{d} t} = \nabla^0 \cdot \mathbb P^{T} + \rho^0 \mathbf g + \eta (\nabla^0)^2 \mathbf v,
\label{momentum-conservation-m}
\end{equation}
where $\eta = \rho^{0} \nu$ denotes the artificial dynamic viscosity.
Following the total Lagrangian formulation, 
the viscous damping term can be discretized as
\begin{equation}
\left( \frac{\text{d} \mathbf v_i}{\text{d} t}\right) ^{\nu} = \eta (\nabla^0)^2 \mathbf v = \frac{2\eta}{m_i}\sum_{j}^{N} V^0_i V^0_j \frac{\mathbf v_{ij}}{r_{ij}} \left( \frac{\partial W_{ij}}{\partial r_{ij}}\right) ^0.
\label{damping-term}
\end{equation}
Note that if the added damping term is solved by an explicit scheme,
the time-step size, along with the acoustic and body-force time-step size criterion
of Eq. (\ref{time-step-criterion}),  
would be also constrained by
\begin{equation}
\Delta t \leq 0.5\frac{h^2}{\nu D},
\label{viscous-criteria-o}
\end{equation}
where $D= \left\lbrace 1, 2, 3 \right\rbrace$ for one-, two- or three-dimensional cases, respectively. 
To achieve fast convergence, 
a suitable large damping ratio is preferred. 
Therefore, 
the overall time-step size may be limited by the above viscous criteria, 
in particular when high spatial resolution is applied, 
and this leads to excessive computational efforts. 

In this work,
we first apply an operator splitting method following Ref. \cite{wang2019split} to decouple the momentum conservation equation Eq.\eqref{momentum-conservation-m} into two procedures, 
i.e. the original momentum part
\begin{equation}
	S_m \quad :\quad \rho^0 \frac{\text{d} \mathbf v}{\text{d} t} = \nabla^0 \cdot \mathbb P^{T} + \rho^0 \mathbf g,
	\label{fist-step-m}
\end{equation}
and the damping part
\begin{equation}
	S_d \quad :\quad \frac{\text{d} \mathbf v}{\text{d} t} = \eta (\nabla^0)^2 \mathbf v.
	\label{second-step-m}
\end{equation}
Here,
two operators $S_m$ and $S_d$ are introduced to represent the original momentum equation without viscosity and the damping process, respectively.
Subsequently,
the first order accuracy Lie-Trotter splitting scheme \cite{mclachlan2002splitting} is applied to approximate the solution from time $t$ to $t+\Delta t$ by
\begin{equation}
	\mathbf v \left( t+\Delta t \right)  =	S_d^{(\Delta t)} \circ	S_m^{(\Delta t)}\mathbf v (t),
	\label{operator-splitting}
\end{equation}
where the symbol $\circ$ denotes the separation of each operator and
indicates that $S_d^{(\Delta t)}$ is applied after $S_m^{(\Delta t)}$.
If implicit methods are adopted to discretize the viscous term,
the damping step can be unconditionally stable and thus a larger time-step size is allowed.

One should note that
large scale matrix operations or iterations over the entire system 
for traditional implicit methods lead to large memory requirement 
and difficulties in paralleization.
To avoid this problem,
we continue to split the entire-domain-related damping step into particle-by-particle operators, e.g. by second-order Strang splitting \cite{strang1968}, as
\begin{equation}
	S_d^{(\Delta t)}  = D_{1}^{(\frac{\Delta t}{2})}  \circ	D_{2}^{(\frac{\Delta t}{2})} \circ \dots \circ D_{N_t-1}^{(\frac{\Delta t}{2})} \circ D_{N_t}^{(\frac{\Delta t}{2})} \circ D_{N_t}^{(\frac{\Delta t}{2})}  \circ	D_{N_t-1}^{(\frac{\Delta t}{2})} \circ \dots \circ D_{2}^{(\frac{\Delta t}{2})} \circ D_{1}^{(\frac{\Delta t}{2})},
	\label{particle-by-particle splitting}
\end{equation}
where $N_t$ denotes the total number of particles and $D_{i}$ the split damping operator corresponding to particle $i$.
In the following, we propose two efficient schemes to the local damping operator, where velocities are updated implicitly and locally.
Then, 
the new time-step velocity updating for the entire field can be achieved 
by carrying out the local split operator to all particles for half a time step and then backwards for another half time step \cite{nguyen2009mass} as shown in Eq.\eqref{particle-by-particle splitting}.

\subsection{Particle-by-particle splitting scheme}\label{subsec:semi-implicit}
In an implicit formulation, 
the local damping term in Eq. \eqref{damping-term} can be rewritten as 
\begin{equation}
\left( \frac{\text{d} \mathbf v_i}{\text{d} t}\right) ^{\nu}  = \frac{2\eta}{m_i}\sum_{j}^{N} V^0_i V^0_j \frac{\mathbf v^{n+1}_{ij} }{r_{ij}} \left( \frac{\partial W_{ij}}{\partial r_{ij}}\right) ^0
\label{damping-term-implicit}
\end{equation}
with $\mathbf v^{n+1}_{ij}  = \mathbf v^{n}_{ij} + \text d \mathbf v_i - \text d \mathbf v_j$.
We denote
\begin{equation}
B_j = 2 \eta  V^0_i V^0_j \frac{1}{r_{ij}} \left( \frac{\partial W_{ij}}{\partial r_{ij}}\right) ^0 \text{d} t
\label{parameters1}
\end{equation}
and 
\begin{equation}
\mathbf E_i = - 2\eta \sum_{j}^{N} V^0_i V^0_j \frac{ \mathbf v^{n}_{ij} }{r_{ij}} \left( \frac{\partial W_{ij}}{\partial r_{ij}}\right) ^0 \text{d} t= - \sum_{j}^{N} B_j \mathbf v^{n}_{ij}.
\label{parameters2}
\end{equation}
The implicit formulation Eq.\eqref{damping-term-implicit} can be further transformed as 
\begin{equation}
\mathbf E_i = \left( \sum_{j}^{N} B_j - m_i\right) \text{d} \mathbf v_i - \sum_{j}^{N} B_j \text{d} \mathbf v_j, 
\label{implicit-equation}
\end{equation}
where $\text{d} \mathbf v_i$ and $\text{d} \mathbf v_j$ represent the incremental change for velocity of particle $i$ and its neighboring particles $j$ induced by viscous acceleration. 
Here, we adopt the gradient descent method \cite{nielsen2015neural} to evaluate $\text{d} \mathbf v_i$ and $\text{d} \mathbf v_j$.  
Specifically, the left hand side (LHS) of Eq. \eqref{implicit-equation} 
can be decreased by following its gradient.
With respect to variables $\left( \text{d} \mathbf v_i, \text{d} \mathbf v_1, \text{d} \mathbf v_2, \cdots, \text{d} \mathbf v_N \right)^T$ ,
the gradient $\nabla \mathbf E_i$ gives
\begin{equation}
\nabla \mathbf E_i = \left( \sum_{j}^{N} B_j - m_i, -B_1, -B_2, \cdots, -B_N \right)^T.
\label{gradient-E}
\end{equation}
We choose 
\begin{equation}
\left( \text{d} \mathbf v_i, \text{d} \mathbf v_1, \text{d} \mathbf v_2, \cdots, \text{d} \mathbf v_N \right)^T = \mathbf k \nabla \mathbf E_i,
\label{gradient-learning-rate}
\end{equation}
where $\mathbf k$ is known as the learning rate \cite{nielsen2015neural}.
By substituting Eqs. \eqref{gradient-E} and \eqref{gradient-learning-rate} into Eq. \eqref{implicit-equation},
the learning rate can be obtained, i.e. 
\begin{equation}
\mathbf k=\left(\left( \sum_{j}^{N} B_j - m_i\right)^2+ \sum_{j}^{N} \left( B_j\right)^2 \right)^{-1} \mathbf E_i. 
\label{learning-rate-solve}
\end{equation}
According to Eqs. \eqref{gradient-E} and \eqref{gradient-learning-rate}, 
the incremental change for velocity by viscous damping can be thus achieved.
In order to ensure momentum conservation,
the velocities of neighboring particles are then modified by the above predicted incremental change.
In summary,
the local update of velocities includes two steps as follows.
The first step calculates the incremental change for velocity by gradient descent method, i.e.
\begin{equation}
\begin{cases}
\mathbf v_i^{n+1}&=\mathbf v_i^{n} + \text{d} \mathbf v_i = \mathbf v_i^{n} + \left( \sum_{j}^{N} B_j - m_i\right) \mathbf k\\
\mathbf v_1^{p}&=\mathbf v_1^{n} + \text{d} \mathbf v_1 = \mathbf v_1^{n}  -B_1 \mathbf k   \\
\mathbf v_2^{p}&=\mathbf v_2^{n} + \text{d} \mathbf v_2 = \mathbf v_2^{n}  -B_2 \mathbf k   \\
&\cdots \\
\mathbf v_N^{p}&=\mathbf v_N^{n} + \text{d} \mathbf v_N = \mathbf v_N^{n}  -B_N \mathbf k   \\
\end{cases},
\label{first-step}
\end{equation} 
where the superscript $p$ denotes the predicted value.
The second step ensures momentum conservation, which yields
\begin{equation}
\begin{cases}
\mathbf v_1^{n+1}&= \mathbf v_1^{n} -B_1 \left( \mathbf v_i^{n+1} -\mathbf v_1^{p} \right) / m_1   \\
\mathbf v_2^{n+1}&= \mathbf v_2^{n} -B_2 \left( \mathbf v_i^{n+1} -\mathbf v_2^{p} \right) / m_2   \\
&\cdots \\
\mathbf v_N^{n+1}&= \mathbf v_N^{n} -B_N \left( \mathbf v_i^{n+1} -\mathbf v_N^{p} \right) / m_N   \\
\end{cases}.
\label{second-step}
\end{equation} 

As the velocities are updated implicitly and much larger time-step size is allowed.
In this work, we choose the following viscous criterion
\begin{equation}
\Delta t \leq 50\frac{h^2}{\nu D}.
\label{viscous-criteria}
\end{equation}
Note that this criterion is about 100 times larger than the corresponding explicit method as presented in Eq. \eqref{viscous-criteria-o}.
Besides, the artificial dynamic viscosity  $\eta = \rho^{0} \nu$ as shown in Eq. \eqref{momentum-conservation-m}
is defined by
\begin{equation}
	\eta = \frac{1}{4} \beta \rho^{0} \sqrt{\frac{E}{\rho}}L=\frac{\beta}{4} \sqrt{\rho^{0} E}L,
	\label{viscous-parameter}
\end{equation}
where $L$ is the characteristic length scale of the problem and $\beta$ denotes a parameter relating to the body shape. Note that choose different value for the parameter $\beta$ 
may variate, though not much, the convergence rate. 
\subsection{Pairwise splitting scheme}\label{subsec:pairwise-splitting}
Inspired by the work of Ref. \cite{litvinov2010splitting},
we present herein the pairwise splitting scheme where particle velocity is updated implicitly and locally in a pairwise fashion. 
By adopting the second-order Strang splitting \cite{strang1968},
the damping operator corresponding to each particle $i$ as given in Eq. \eqref{particle-by-particle splitting} 
is further split based on its neighbors, i.e.
\begin{equation}
	D_i^{(\Delta t)}  = D_{i,1}^{(\frac{\Delta t}{2})}  \circ	D_{i,2}^{(\frac{\Delta t}{2})}  \circ \dots \circ  D_{i,N-1}^{(\frac{\Delta t}{2})} \circ  D_{i,N}^{(\frac{\Delta t}{2})} \circ D_{i,N}^{(\frac{\Delta t}{2})}  \circ	D_{i,N-1}^{(\frac{\Delta t}{2})}  \circ \dots \circ  D_{i,2}^{(\frac{\Delta t}{2})} \circ  D_{i,1}^{(\frac{\Delta t}{2})},
	\label{pairwise-splitting}
\end{equation}
where $D_{i,j}^{(\frac{\Delta t}{2})}$ denotes the interaction between particle $i$ and its neighbors.
Specifically,
the incremental changes for velocity of a specific particle pair induced by viscosity can be written in implicit form as
\begin{equation}
	\begin{cases}
		m_i \text{d} \mathbf v_i =  B_{j} \left( \mathbf v_{ij} + \text d \mathbf v_i - \text d \mathbf v_j\right) \\
		m_j \text{d} \mathbf v_j = -B_{j} \left( \mathbf v_{ij} + \text d \mathbf v_i - \text d \mathbf v_j\right) \\
	\end{cases}.
	\label{pairwise-eq}
\end{equation} 
Here, 
$B_j$ is defined in Eq. \eqref{parameters1} and it is obvious that this process does not change the conservation of momentum.
Then, $\text{d} \mathbf v_i$ and $\text{d} \mathbf v_j$ can be obtained straightforwardly by solving Eq. \eqref{pairwise-eq},
which yields
\begin{equation}
	\begin{cases}
		\text{d} \mathbf v_i = m_j\frac{  B_{j} \mathbf v_{ij}}{ m_i m_j - \left( m_i + m_j\right) B_j } \\
		\text{d} \mathbf v_j = - m_i\frac{ B_{j} \mathbf v_{ij}}{ m_i m_j - \left( m_i + m_j\right) B_j } \\
	\end{cases}.
	\label{pairwise-solve}
\end{equation}
By sweeping over all neighboring particle pairs for half a time step and then backwards for another half time step,
the incremental changes for velocity of particle $i$ and all its neighbors can be thus achieved.
Compared to the particle-by-particle splitting method in Section \ref{subsec:semi-implicit},
this scheme leads more errors in solving viscosity due to the further splitting in pairwise fashion.
However, it is unconditional stable and thus more suitable for problems with high spatial resolution and high damping ratio. 
\subsection{Random-choice strategy}\label{subsec:random-chosen}
As noted in Ref.\cite{lin2014efficient},
the solutions for nonlinear problems are path dependent 
and may vary with different damping ratio.
The added viscous force would hinder the system achieving correct static state 
especially when the damping radio is large.
Thus, a suitable damping ratio has to be selected \cite{lin2014efficient,crisfield1997non}
for faster convergence of the system with the aforementioned viscous damping methods.

To avoid this problem and relax the limitation on the choice of large damping ratio,
we present herein a random-choice strategy 
in which the viscosity term is imposed randomly rather than at every time step.
For this, the artificial dynamic viscosity $\eta$ is modified as
\begin{equation}
\tilde{\eta} = 
\begin{cases}
\eta / \alpha  \quad \quad & {\rm if} \quad \alpha < \phi\\
0      & {\rm otherswise}   \\
\end{cases},
\label{viscosity-reset}
\end{equation}
where $\phi$ is a random number uniformly distributed between $0$ to $1$.
$\alpha$ is a parameter determining the probability, 
and we set $\alpha= 0.2$ in this work.
Therefore, the resistance on displacement 
induced by the large artificial viscosity can be released randomly,
as will shown the numerical example (Section \ref{bending-cantilever}), 
which eliminates the path dependency and accelerates the convergence.
Note that this strategy also helps to save much of computational cost 
since the computation of damping is only carried out 
at a small fraction of time steps.
\subsection{Splitting cell-linked list }\label{subsec:splitting cell-linked list}
Shared-memory parallel technique is widely used in SPH simulation due to its easy-implementation nature. 
However, 
for the above two proposed implicit approaches,
implementing shared-memory approach taking advantage of multicore performance, is problematic.
Since the velocities of particle $i$ and the neighbors are updated simultaneously in above implicit method, 
conflicts may happen when the values of one particle pair are updated at the same time by multiple threads.

\begin{figure}[htb!]
	\centering
	\includegraphics[trim = 2cm 4cm 2cm 3.8cm, clip,width=0.90\textwidth]{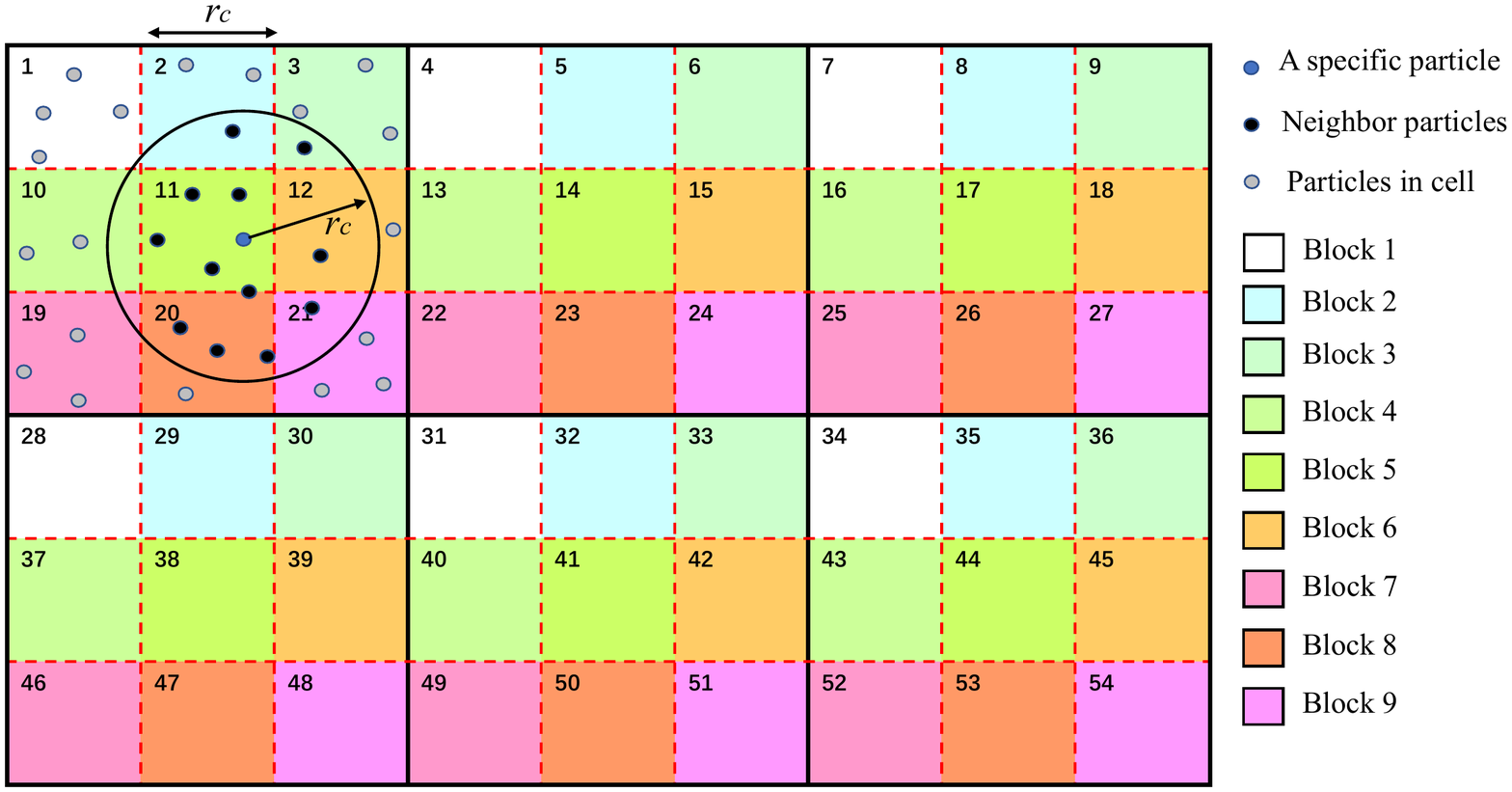}
	\caption{Sketch of the splitting CLL method in two-dimensional space.
	}
	\label{split-cell-linked}
\end{figure}

In order to avoid this problem, a splitting CLL method is proposed.
Figure \ref{split-cell-linked} presents a sketch of the splitting CLL method in two-dimensional space.
Similar to the conventional CLL method,
the computational domain is divided into cells with the same size to reduce the searching space \cite{liu2003smoothed}
and the cell spacing $r_c$ is equal to the cut-off radius of the kernel function. 
Then, 
the cell domain is divided into $9$ blocks ($27$ blocks for three-dimensional space) and all the cells are assigned to blocks. 
To implement the share-memory parallelization between blocks, 
each block satisfies the condition that all its cells are not neighboring cells to each other. 
As shown in Figure \ref{split-cell-linked}, 
cells are assigned to block $1$ with condition  of  $\left( \text{Cell}_i  \subset \text{Block}1 \middle| \left\lbrace i =  1,4,7,28,31,34  \right\rbrace\right) $ 
and also cells are assigned to block $9$ with condition  of  $\left( \text{Cell}_i  \subset \text{Block}9  \middle| \left\lbrace i =  21,24,27,48,51,54 \right\rbrace\right) $.
When the tasks are distributed to threads based on blocks,
the particles executed by different threads are at least two cell space away from each other at any time.
The conflicts can be thus avoid, since the neighboring particles only locate in the same cell or other adjacent 8 cells.
%
%
%
\section{Numerical examples}\label{validation}
In this section, 
various tests for static equilibrium simulation including  
cantilever bending and twisting, ball free falling and a fluid-structure interaction problem
are simulated to demonstrate the effectiveness and efficiency of the present method. 
In the following cases, the 5th-order C2 Wendland kernel \cite{wendland1995} is adopted and the smoothing length $h=1.3dp$ with $dp$ denoting the initial particle spacing.
Term "No damping" denotes the conventional TLSPH method without damping imposed,
term "Reference" is the static solution obtained by the conventional TLSPH method and 
term "Present" represents the particle-by-particle splitting damping method presented in Section \ref{subsec:semi-implicit} if not mentioned otherwise.
\subsection{Bending cantilever}\label{bending-cantilever}
\begin{figure}[htb!]
	\centering
	\includegraphics[trim = 10cm 10cm 10cm 7cm, clip,width=0.6\textwidth]{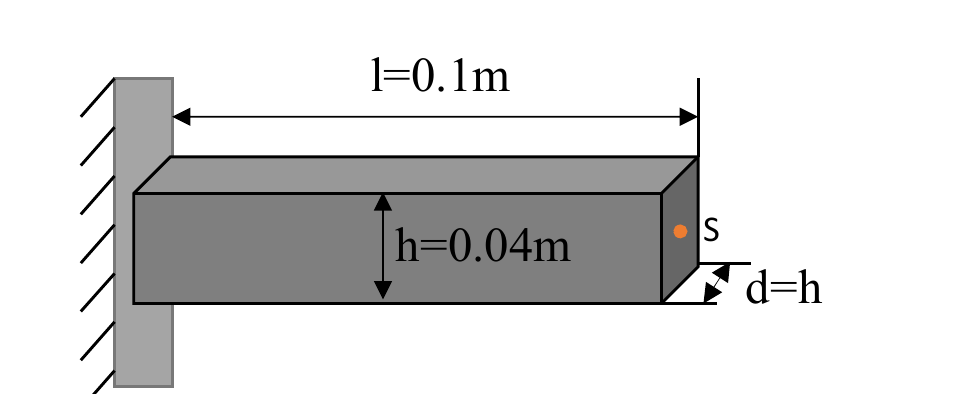}
	\caption{Bending cantilever: Initial configuration. Probe $S$ is located at the free-end of the cantilever. }
	\label{figs:bending-cantilever-setup}
\end{figure}

We first consider a three dimensional cantilever bending under gravity. 
As shown in Figure \ref{figs:bending-cantilever-setup},
one end of the cantilever is clamped to the wall and the body bends freely under the gravity $g=9.8 m/s^2$.
The neo-Hookean material with density $\rho=1265 kg/m^3$, Young's modulus $E=5\times 10^4 Pa$ and Poisson's ratio $\nu=0.45$ is applied.
This case is studied with three different spatial resolutions, i.e. $h/dp = \left\lbrace 6, 12, 24 \right\rbrace $.
The artificial viscosity $\eta_0=\frac{\beta}{4}\sqrt{\rho E} d = 32 kg/(m\cdot s)$ with $\beta=d/l=0.4$ is applied. 

\begin{figure}[htb!]
	\centering
	\includegraphics[trim = 1cm 2cm 1cm 2cm, clip,width=.95\textwidth]{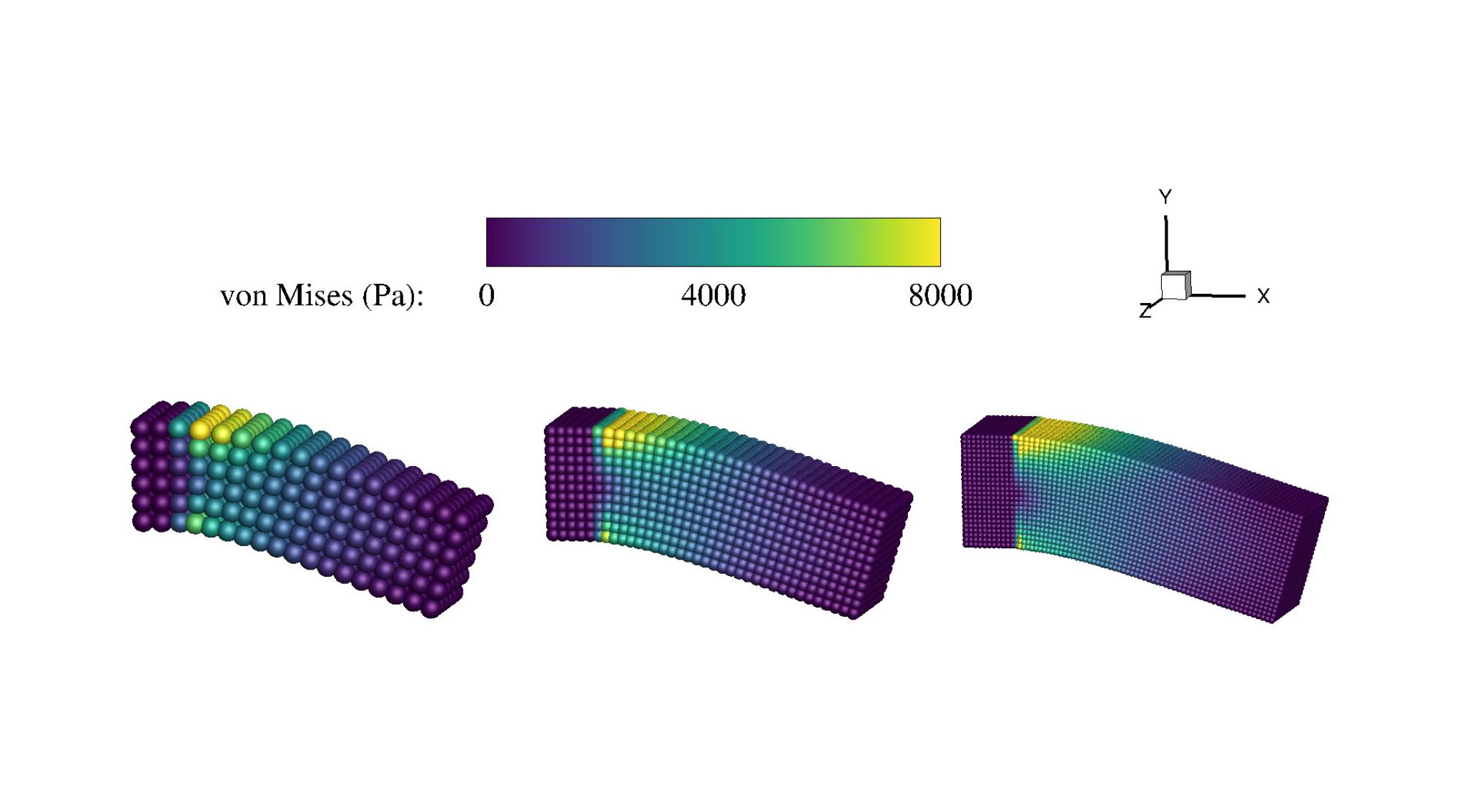}
	\caption{Bending cantilever: Deformed configuration in steady state at resolution of $h/dp = 6$ (left panel), $h/dp = 12$ (middle panel)
		and $h/dp = 24$ (right panel).
	}
	\label{figs:bending-cantilever-configuration}
\end{figure}

\begin{figure}[htb!]
	\centering
	\includegraphics[trim = 0cm 0.2cm 0cm 0cm, clip,width=.8\textwidth]{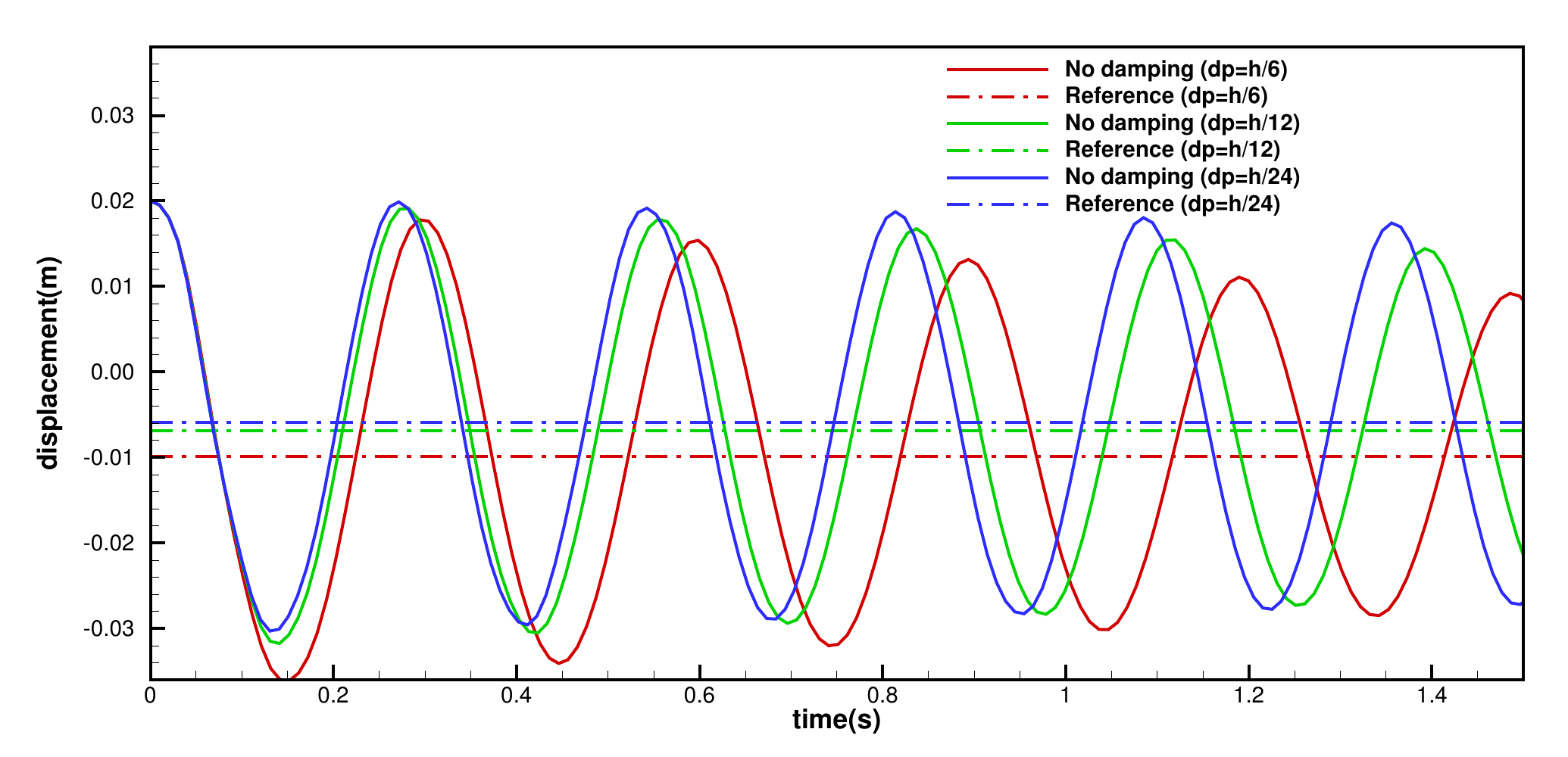} \\
	\includegraphics[trim = 0cm 0.2cm 0cm 0cm, clip,width=.8\textwidth]{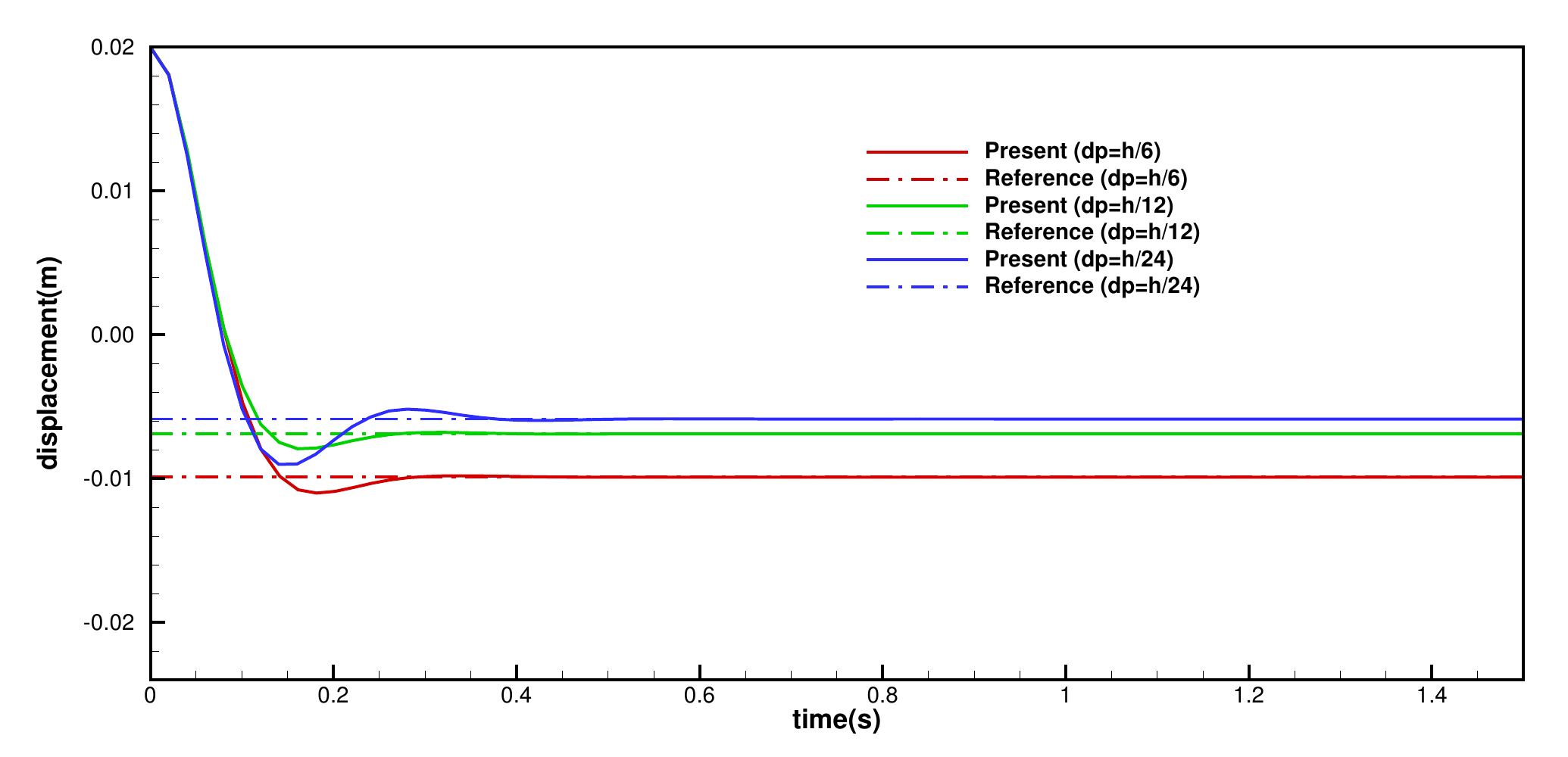} 
	\caption{Bending cantilever: The time history displacement of the end point $S$ in vertical direction obtained by conventional method (top panel) and the present method (bottom panel).	}
	\label{figs:bending-cantilever-three-resolution}
\end{figure}

Figure \ref{figs:bending-cantilever-configuration} presents the deformed configuration in equilibrium state with three different spatial resolutions.
The time histories of the displacement in $y-$direction of the point $S$ are given in Figure \ref{figs:bending-cantilever-three-resolution}.
It is clear that the conventional TLSPH method exhibits oscillations and takes long time to reach the steady state. 
As for the present method, only a few oscillations are exhibited and the steady state is reached quickly which reduces much computational time for static solution.
Besides, with the conventional TLSPH method, 
the final displacements of point $S$ in $y-$direction with $h/dp = 6$, $h/dp = 12$ and $h/dp = 24$ 
are $-9.90 \times 10^{-3} m$, $-6.89 \times 10^{-3} m$ and $-5.86 \times 10^{-3} m$, respectively.
We can observe that the same static results are obtained by the present method and  
a converge solution is achieved with increasing resolutions. 

\begin{figure}[htb!]
	\centering
	\includegraphics[trim = 0cm 0.2cm 0cm 0cm, clip,width=.8\textwidth]{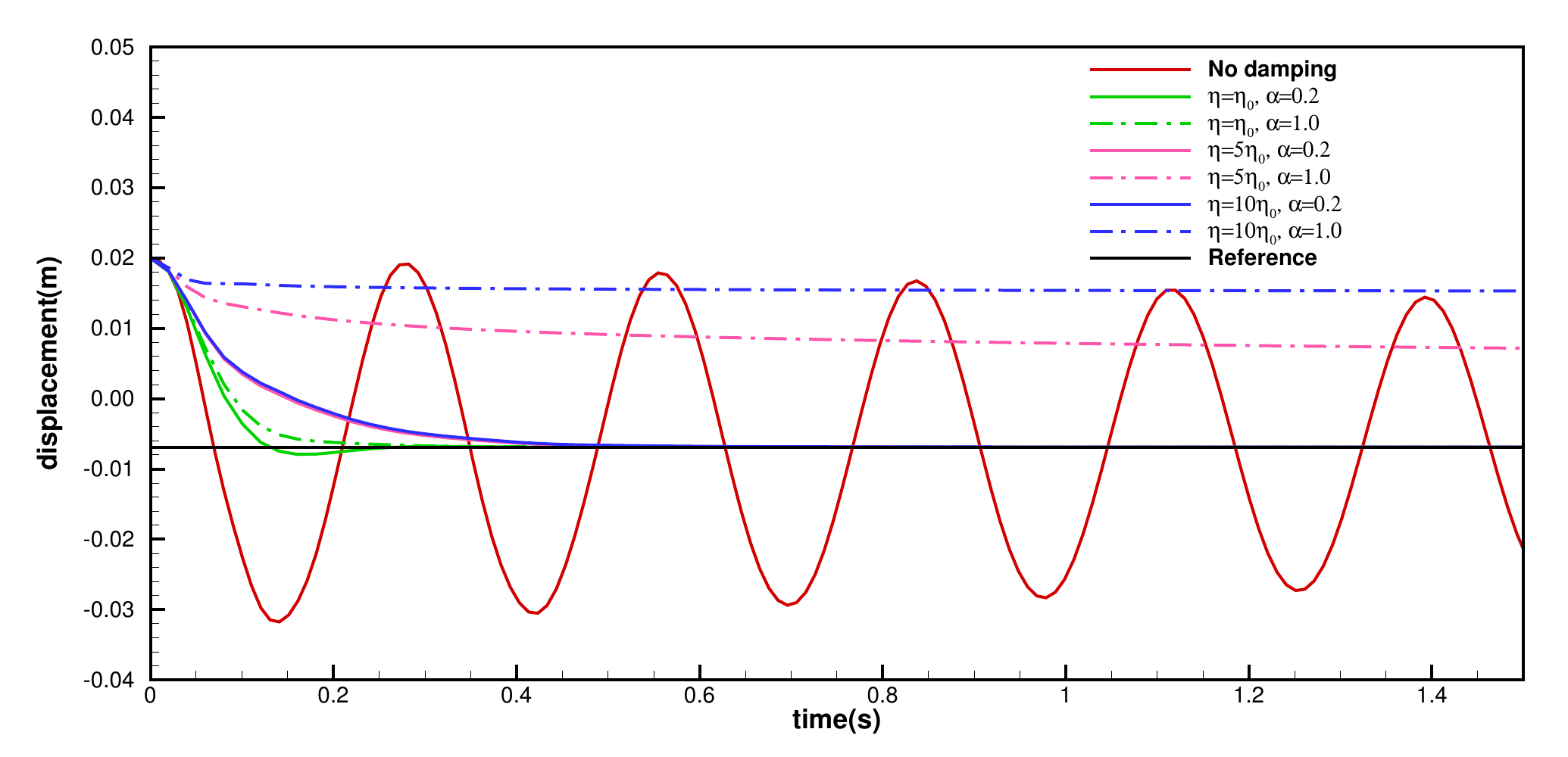} \\
	\includegraphics[trim = 0cm 0.2cm 0cm 0cm, clip,width=.8\textwidth]{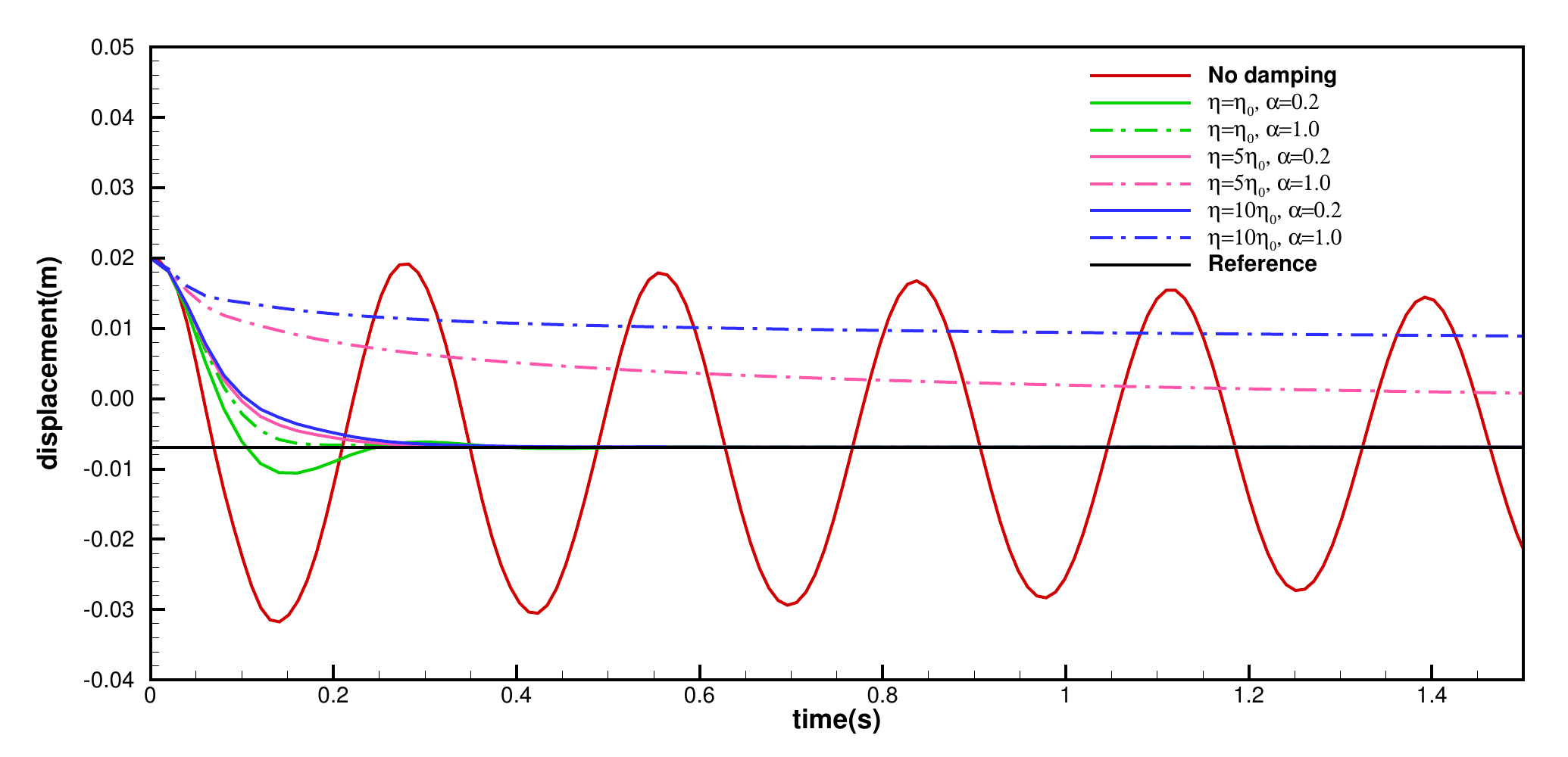}
	\caption{Bending cantilever: Comparisons of the time history displacement of point $S$ at the resolution of $h/dp = 12$ by the operator splitting method (top panel) and the pairwise splitting method (bottom panel).
	}
	\label{figs:bending-cantilever-compare}
\end{figure}

To test the performance of the adopted random-choice strategy,
the setup with parameter $\alpha=1$ of Eq. \eqref{viscosity-reset} and artificial viscosity $\eta=5\eta_0$ and $\eta=10\eta_0$ are also considered.
The top panel of Figure \ref{figs:bending-cantilever-compare} shows the comparison of the present method with different parameters at the resolution of $h/dp=12$.
We can observe that the results of $\alpha=1$, in which the viscous damping is imposed at every time-step, 
cannot achieve the final state even at $t=1.5s$ especially for higher artificial viscosity.
Also note that lager damping leads larger distance between the solution to that of the final state and the velocity approaching to final state becomes very small due to too much damping.
However, with the help of the random-choice strategy, e.g. $\alpha=0.2$, 
the system is released randomly from large damping and the same final state can be reached efficiently for both $\eta=5\eta_0$ and $\eta=10\eta_0$.
Besides, the pairwise splitting scheme as presented in Section \ref{subsec:pairwise-splitting} is also introduced to discretize the viscous term in this case.
The bottom panel of Figure \ref{figs:bending-cantilever-compare} gives the corresponding results with different parameters at the resolution of $h/dp=12$.
Compared to that of the top panel, 
it can be observed that the viscosity achieved by the pairwise splitting scheme 
is less than the splitting scheme presented in Section \ref{subsec:semi-implicit}.
However, with higher artificial viscosity, the steady state of the cantilever can also be achieved efficiently.
Meanwhile, 
this pairwise splitting scheme is more stable and free from the limitation of viscous criterion, which is more suitable for the problem with high resolution and high damping ratio.

The computational performance of the present method by implementing the shared-memory parallelization and random-choice strategies is then tested.
The computation is conducted with spatial resolution of $h/dp=12$ and the total physical time of $t = 2s$ on an AMD Ryzen 5 3600 6-Core CPU Desktop computer with 16GB RAM and Windows system (10).
Table \ref{bending-cantilever-time} reports the CPU wall-clock time for the process of viscous damping.
We can see that the shared-memory parallelization strategy demonstrates about 2.8 speedup compared to that of serial computing.
Besides, the random-choice strategy also saves much computational efforts and only takes $20\%$ ($\alpha=0.2$) CPU time compared to the one without applying this strategy.
The viscous damping process only takes $2.38s$ during the whole computation and
optimized computational performance is achieved.

\begin{table}
	\scriptsize
	\centering
	\caption{CPU time for viscous damping process in the simulation of Bending cantilever with different parameters.}
	\begin{tabularx}{8.5cm}{@{\extracolsep{\fill}}lcc}
		\hline
		CPU time (s) & serial computation & parallel computation\\
		\hline
		$\alpha=1.0$ & $32.82$	& $11.61$  \\
		\hline
		$\alpha=0.2$ & $6.66$	& $2.38$  \\
		\hline
	\end{tabularx}
	\label{bending-cantilever-time}
\end{table}
%
\subsection{Twisting cantilever}
\label{twisting-cantilever}
\begin{figure}[htb!]
	\centering
	\includegraphics[trim = 9cm 7cm 9cm 5cm, clip,width=0.6\textwidth]{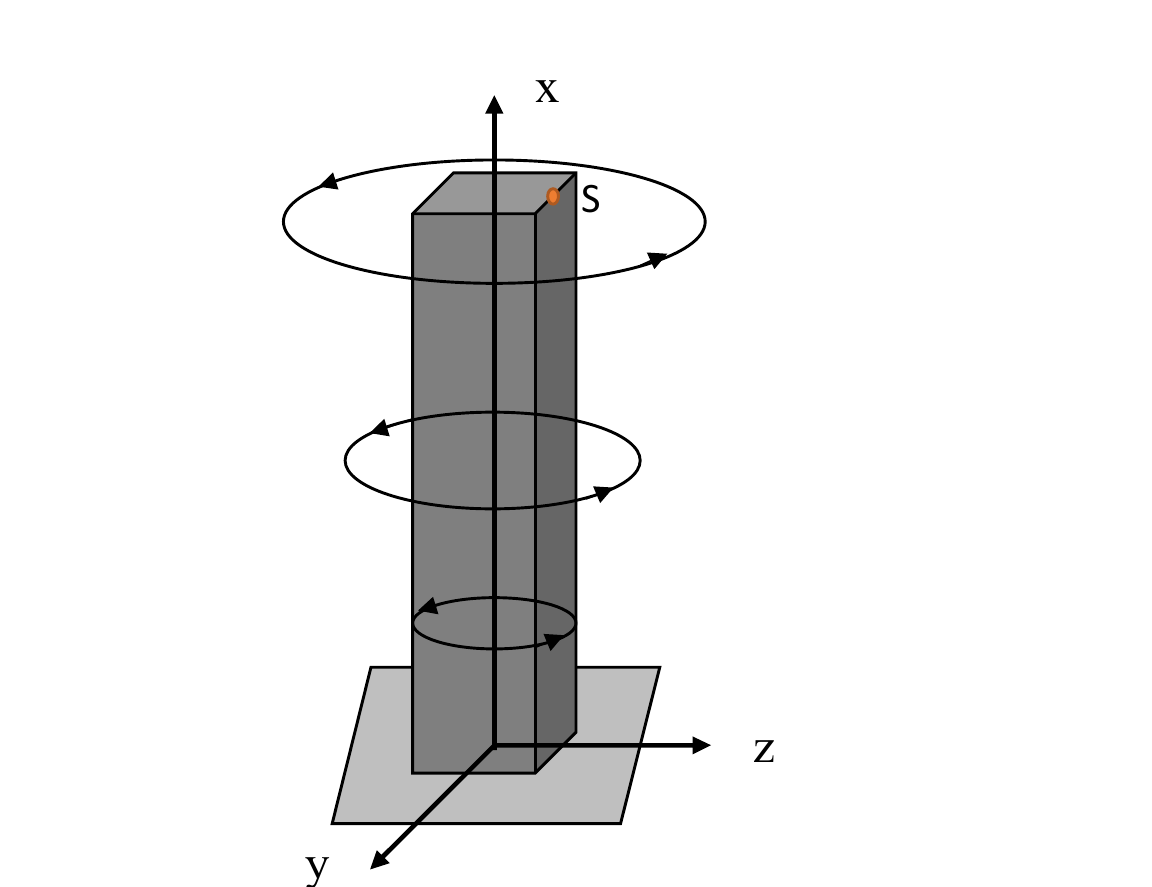}
	\caption{Twisting cantilever: Initial configuration. Probe $S$ locates at the free-end of the cantilever. }
	\label{figs:twisting-cantilever-setup}
\end{figure}

Under a rotational body-force, the cantilever as presented in Section \ref{bending-cantilever} can be twisted from the initial configuration.
As shown in Figure \ref{figs:twisting-cantilever-setup},
the bottom of the cantilever is also clamped and a sinusoidal rotational body-force is imposed.
The body-force relates to the origin and is defined by
\begin{equation}
	g = \left[0,y \gamma, z \gamma \right] ^{T}, 
	\label{rotation-body-force}
\end{equation}
where $\gamma = \frac{20g}{h} \text{sin}(\pi x /2L)$ and $g=9.8 m/s^2$.
The similar neo-Hookean material property as in bending case is adopted and this case is performed at the resolution of $h/dp=12$.

\begin{figure}[htb!]
	\centering
	\includegraphics[trim = 0cm 0.2cm 0cm 0cm, clip,width=.8\textwidth]{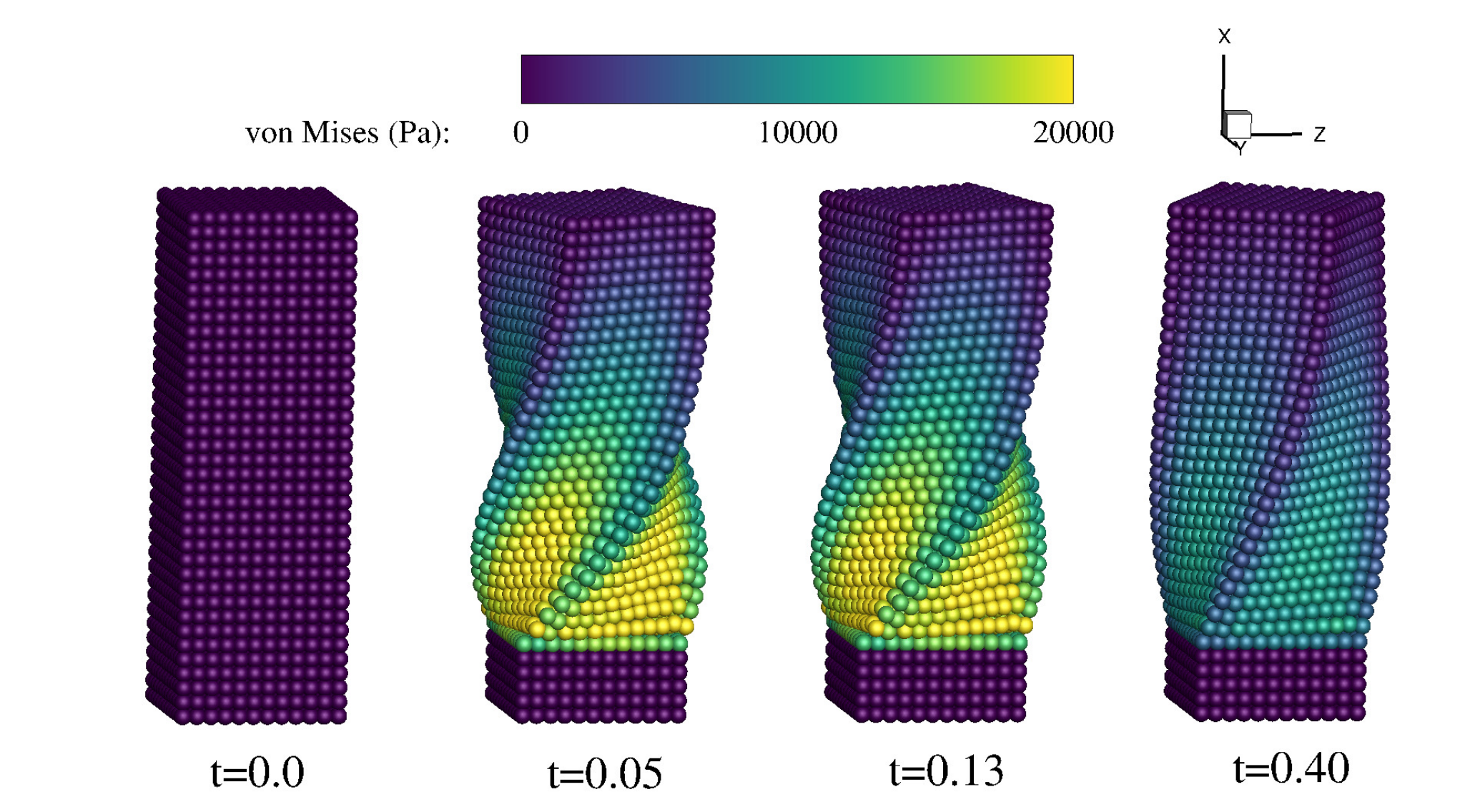} 
	\caption{Twisting cantilever: Four time instants as labeled by blue dots in Figure \ref{figs:twisting-cantilever-compare} of deformation and von Mises stress.
	}
	\label{figs:twisting-cantilever-contour}
\end{figure}

\begin{figure}[htb!]
	\centering
	\includegraphics[trim = 0cm 0.2cm 0cm 0cm, clip,width=.8\textwidth]{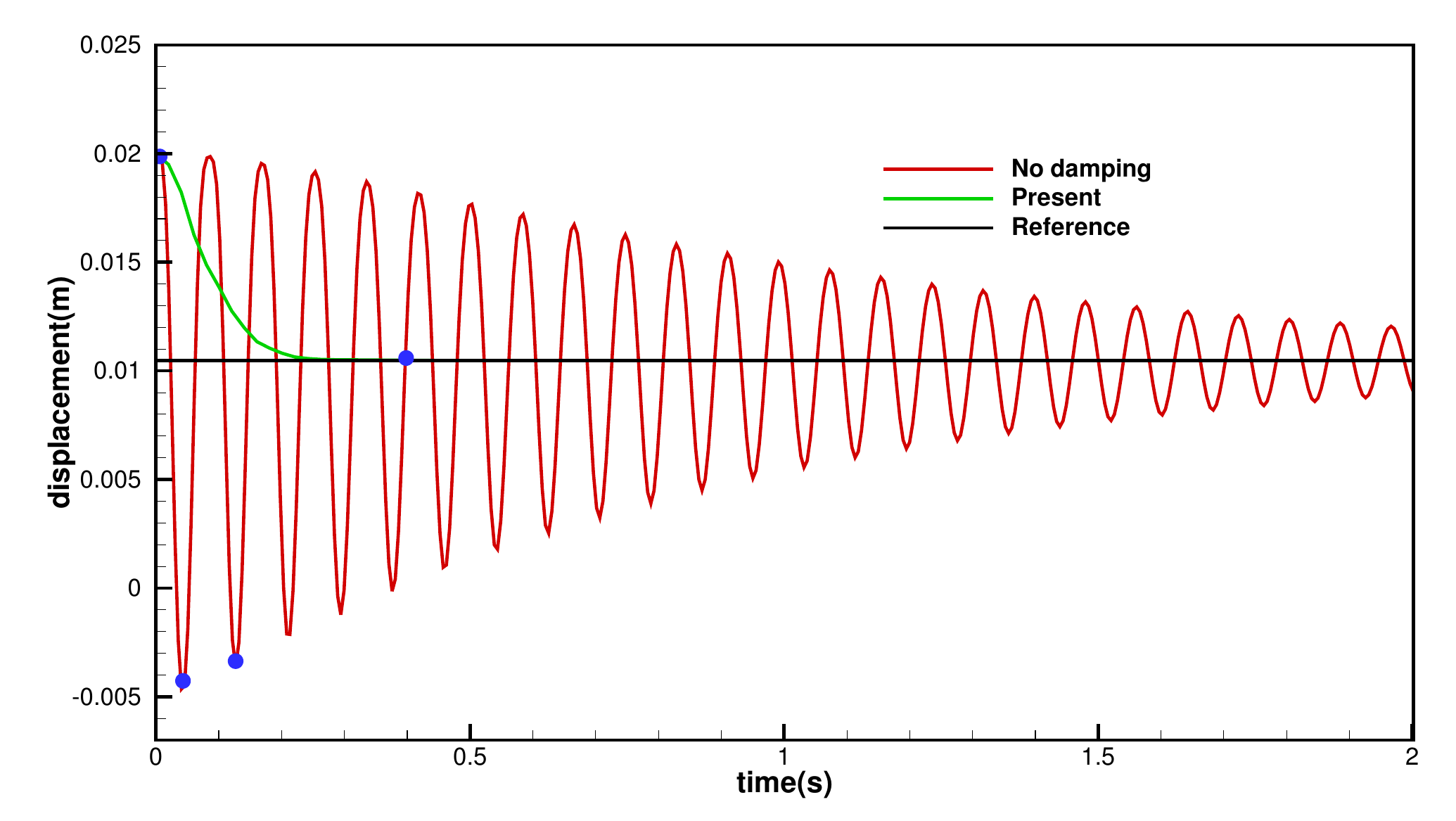}
	\caption{Twisting cantilever: The time history of displacement at the end point $S$ in $z-$ direction.	}
	\label{figs:twisting-cantilever-compare}
\end{figure}

Figure \ref{figs:twisting-cantilever-contour} and \ref{figs:twisting-cantilever-compare}
present several snapshots of the deformation and the time history displacement in $z-$direction of the point $S$, respectively.
With the present viscous damping imposed, 
the final steady state of the system is reached much faster and the same result with that of the conventional TLSPH is achieved
implying the efficiency and accuracy of the present method.

%
\subsection{Ball free falling}\label{subsec0}
\begin{figure}[htb!]
	\centering
	\includegraphics[trim = 11cm 7cm 11cm 6cm, clip,width=0.45\textwidth]{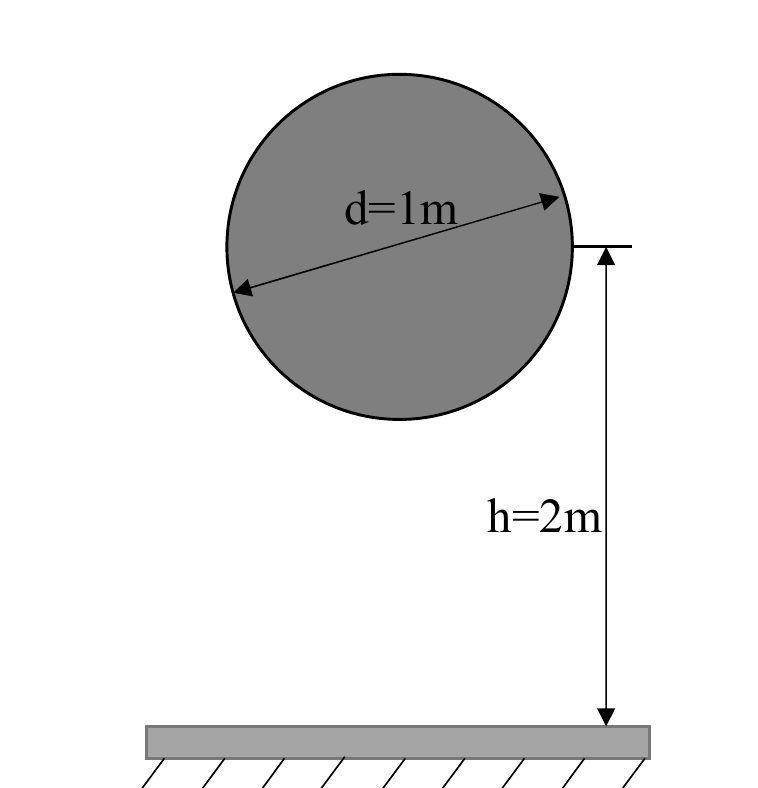}
	\includegraphics[trim = 1.0cm 1.0cm 1.0cm 1.5cm, clip,width=0.45\textwidth]{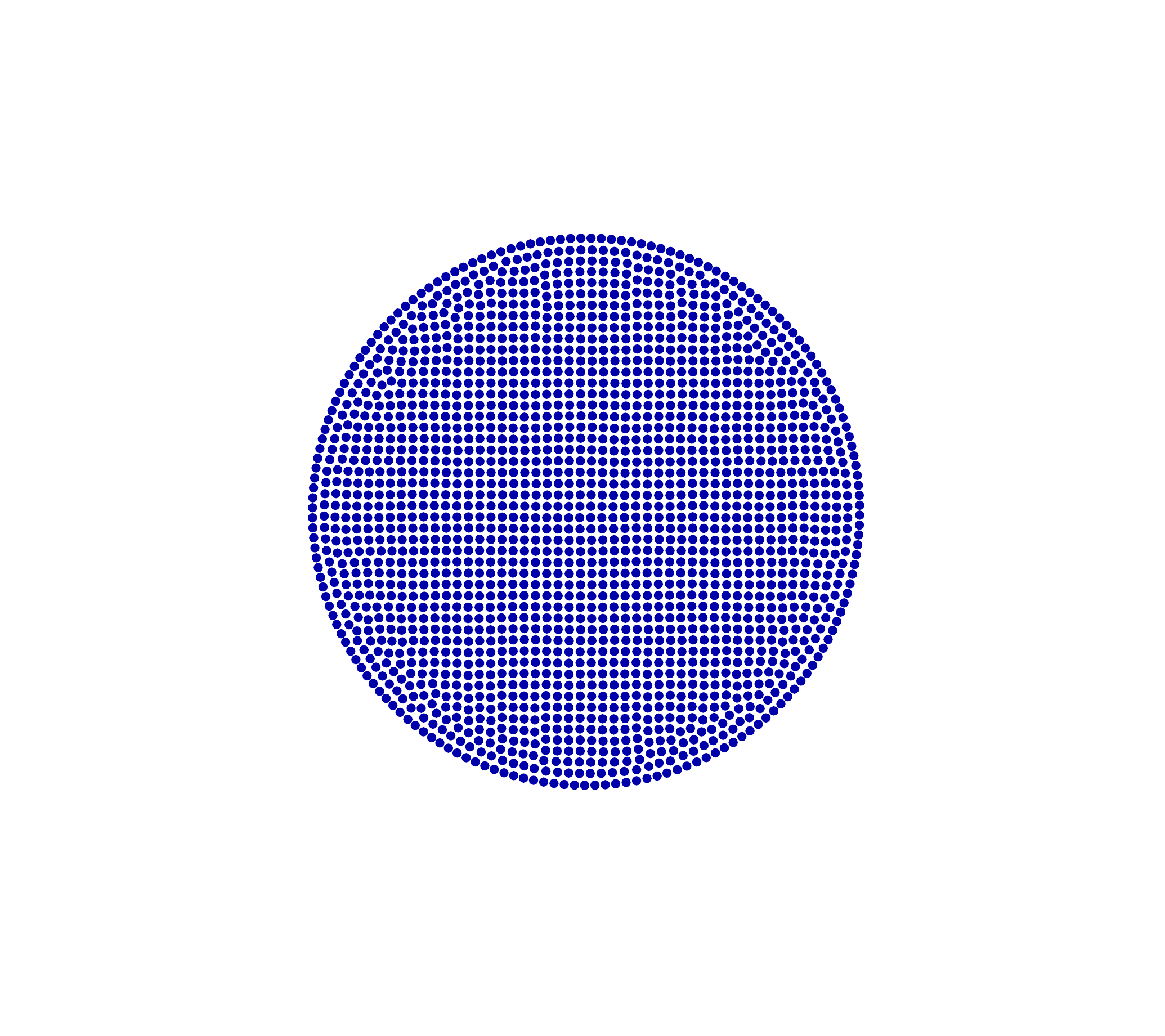}
	\caption{Ball free falling: Initial configuration (left panel) and body-fitted particle distribution of the ball (right panel). }
	\label{figs:ball-collision-setup}
\end{figure}

A two dimensional ball free falling is considered herein. 
Figure \ref{figs:ball-collision-setup} gives the initial configuration and the body-fitted particle distribution of the ball.
In this case, a ball falls freely accelerated by gravity $g=9.8 m/s^2$ and collides with the floor.
A linear elastic material with density $\rho=1000 kg/m^3$, Young's modulus $E=5\times 10^5 Pa$ and Poisson's ratio $\nu=0.45$ is adopted 
and the initial particle spacing $dp = 0.02d$. 
For the present method, 
the artificial viscosity is given by $\eta=\frac{\beta}{4}\sqrt{\rho E} d = 5590 kg/(m\cdot s)$ with $\beta = 1$.

\begin{figure}[htb!]
	\centering
	\includegraphics[trim = 0cm 2cm 0cm 2cm, clip,width=.95\textwidth]{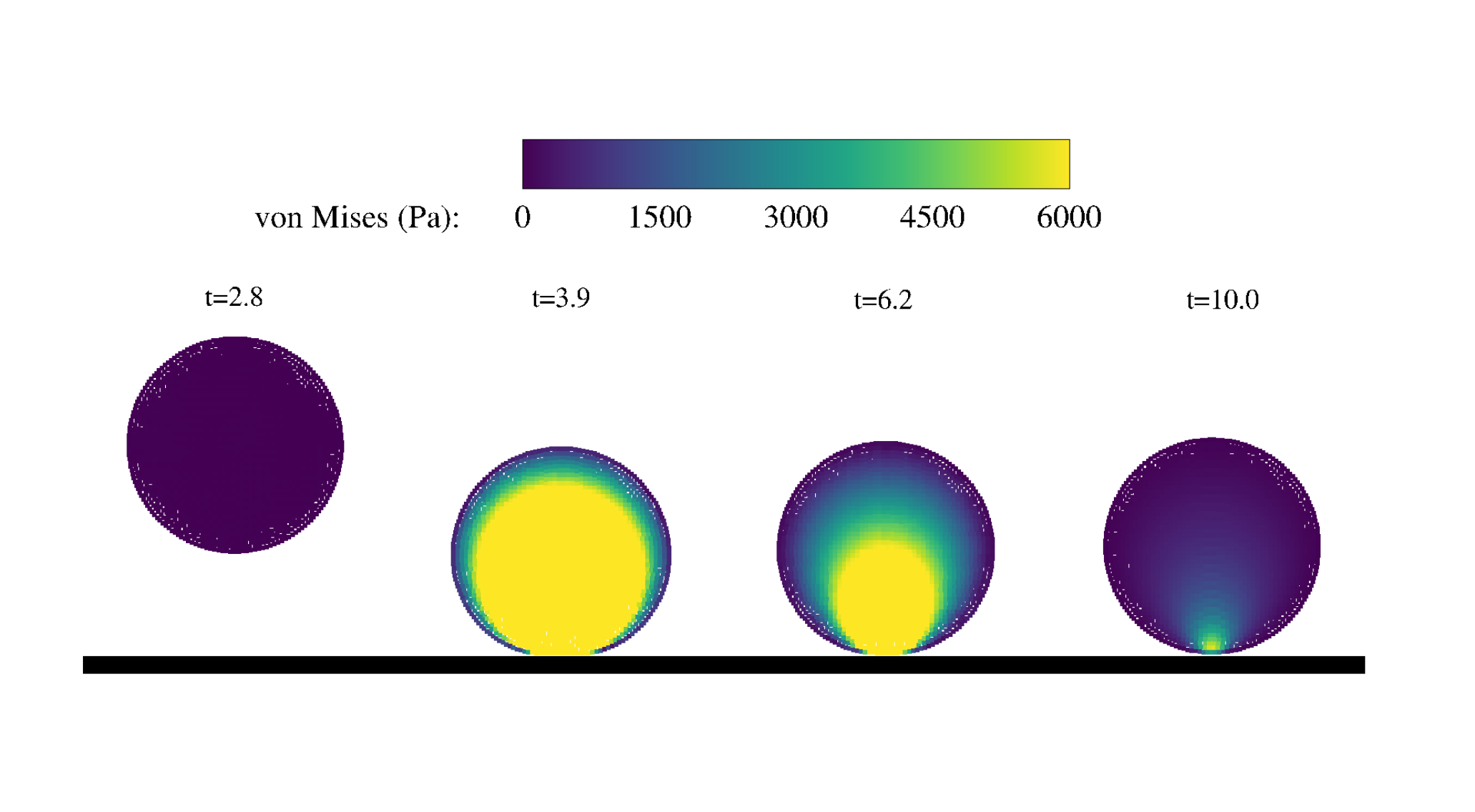}
	\caption{Ball free falling: Four time instants as labeled by blue dots in Figure \ref{figs:ball-collision-compare} of the ball deformation and von Mises stress.
	}
	\label{figs:ball-collision-instant}
\end{figure}

\begin{figure}[htb!]
	\centering
	\includegraphics[trim = 0cm 0.2cm 0cm 0.2cm, clip,width=0.8\textwidth]{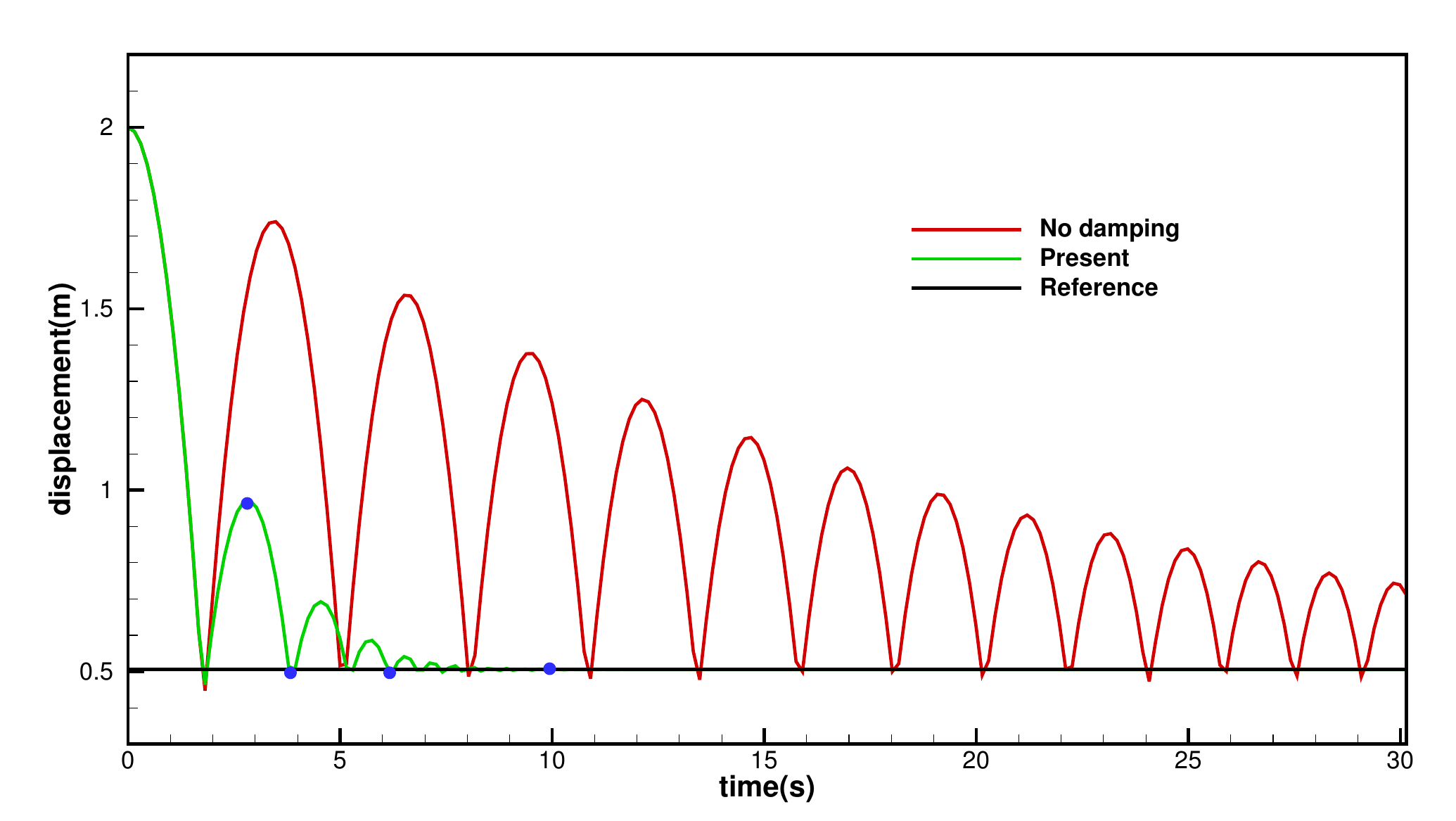}
	\caption{Ball free falling: The time history of the displacement at ball center in vertical direction. }
	\label{figs:ball-collision-compare}
\end{figure}

Figure \ref{figs:ball-collision-instant} shows four snapshots of the ball deformation and the corresponding von Mises stress.
After several collisions, the equilibrium of this system is reached and the ball stands still on the wall. 
Figure \ref{figs:ball-collision-compare} gives the time history of the displacement of the ball center in vertical direction.
It is obvious that achieving steady state of the system for the conventional TLSPH method is time-consuming and more collisions are exhibited in the simulation.
The amplitude reduction is mainly caused by the numerical dissipation of the scheme.
However, with the help of the present viscous damping method,
the system achieves steady state faster and only few collisions is exhibited.
Term "Reference" in Figure \ref{figs:ball-collision-compare} denotes the final displacement obtained by the conventional TLSPH method and
it is clear that the present method achieves the same final results with that of the classic method.
%
\subsection{Hydrostatic water column on an elastic plate}
\label{subsec4}
\begin{figure}[htb!]
	\centering
	\includegraphics[trim = 4cm 4cm 4cm 4cm, clip,width=0.6\textwidth]{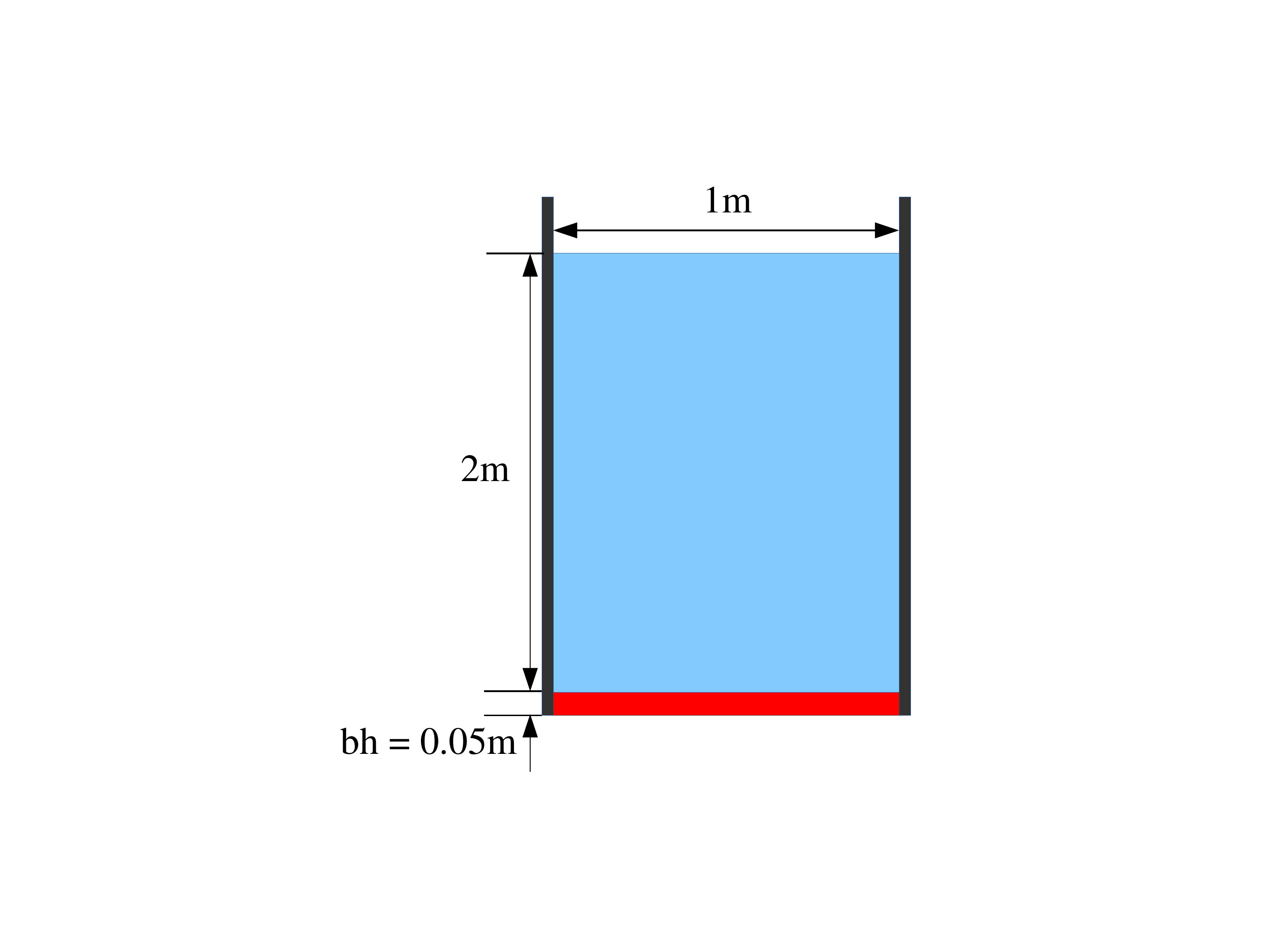}
	\caption{Hydrostatic water column on an elastic plate: Initial configuration. }
	\label{figs:hydrostatic-setup}
\end{figure}

This case is taken from Ref.\cite{zhang2020multi} and relates to the deformation of an elastic plate induced by the hydrostatic pressure of a water column.
The geometric parameters and basic setups are given in Figure \ref{figs:hydrostatic-setup}.
An aluminium plate with thick $bh=0.05m$ is suddenly exposed to the hydrostatic pressure loading by a $2m$ height water column and
an equilibrium state can be reached after initial oscillations.
Following Refs. \cite{zhang2020multi, fourey2017efficient, khayyer2018enhanced},
the material with density of $\rho=2700 kg/m^3$, Young's modulus $E=67.5GPa$ and Poisson's ratio $0.34$ is applied for the aluminium plate. 
As for the water,
the density is $\rho=1000 kg/m^3$ and the weakly-compressible model is adopted with
the sound speed of $c=10 U_{max}$,
where $U_{max}= 2\sqrt{gH} =8.86 m/s$ is the maximum anticipated flow speed.
In fluid-structure coupling, multiple time steps are applied.
Similar to  Refs. \cite{zhang2020multi, fourey2017efficient, khayyer2018enhanced}, 
a constant fluid time-step size $\Delta t^{F}=2.0\times 10^{-5} s$ is adopted.
More details of fluid structure interaction formulation with SPH method are referred to Ref. \cite{zhang2020multi}.
In this case, the viscous damping is also introduced to water part 
and the corresponding artificial dynamic viscosity is given by $\eta_f= \rho U_{max} L_f = 8858.9 kg/(m\cdot s)$
with $L_f$ denoting the characteristic length.
According to theoretical solution,
the magnitude of the static deformation at the mid-span of the plate is $d=-6.85 \times 10 ^{-5} m$.

\begin{figure}[htb!]
	\centering
	\includegraphics[trim = 0cm 0.5cm 0cm 0cm, clip,width=.8\textwidth]{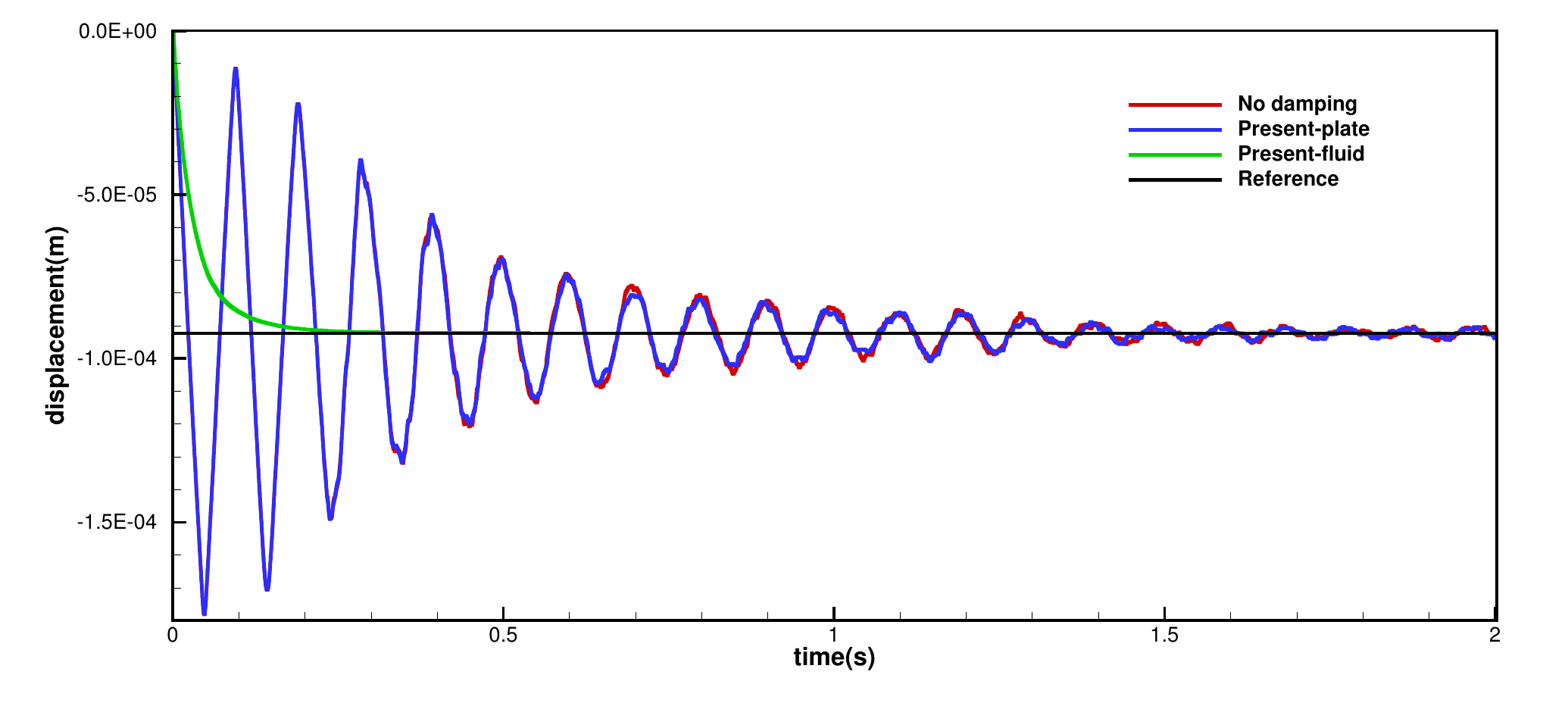} \\
	\includegraphics[trim = 0cm 0.5cm 0cm 0cm, clip,width=.8\textwidth]{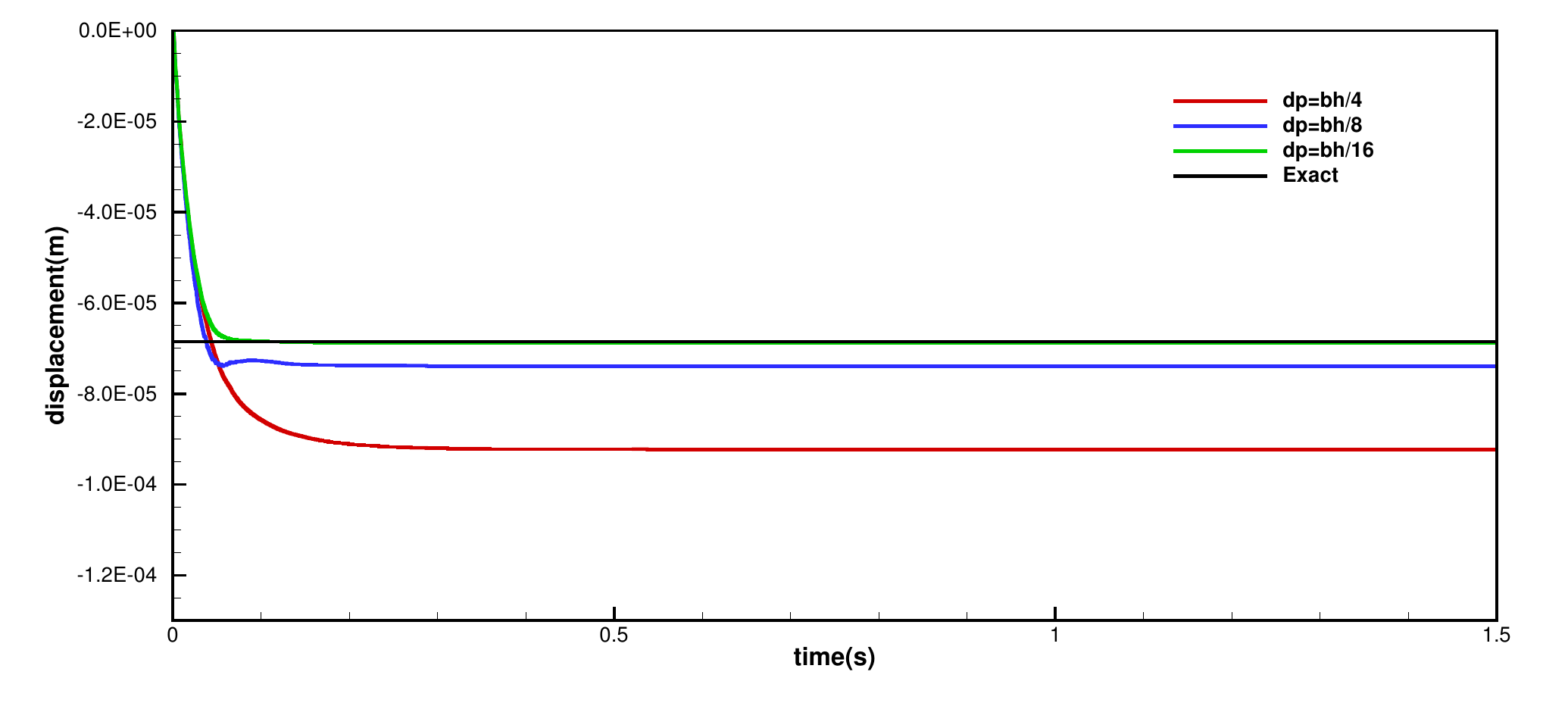} 
	\caption{Hydrostatic water column on an elastic plate: Mid-span displacement at resolution of $dp = bh/4$ (top panel) and a convergence study with increasing spatial resolutions (bottom panel).	}
	\label{figs:hydrostatic}
\end{figure}

The time history of mid-span displacement for spatial resolution $bh/dp = 4$ is presented in the top panel of Figure \ref{figs:hydrostatic}.
Term "Present-plate" denotes the viscous damping only imposed on the elastic plate, 
"Present-fluid" is the solution with viscous damping introduced to water part and "Exact" represents the theoretical solution.
It can be observed that the solution with plate damping is similar to that of without artificial damping imposed and oscillations are not repressed effectively. 
In fact, the oscillation of the plate is mainly induced by the water. 
When the present viscous damping is introduced to water, the solution reaches the equilibrium state rapidly as shown in Figure \ref{figs:hydrostatic}.
Moreover, a convergence study with the present viscous damping imposed on water part is carried out by increasing the spatial resolution.
The results are given in bottom panel of Figure\ref{figs:hydrostatic} and it is clear that a convergent displacement of the mid-span is achieved implying the accuracy of the present method.

The CPU wall-clock time of the viscous damping process is recorded in Table \ref{hydrostatic-time} until the physical time of $t=2s$.
It can be observed that the damping process only takes $26.89s$ during the computing and almost $5$ times speedup is achieved by implementing share-memory parallelization. 
Besides, if an explicit scheme instead of the present implicit schemes is implemented for viscous damping,
the fluid time-step size would be limited to $\Delta t^F \leq 0.25\frac{h^2}{\nu}=4.4 \times 10^{-6}s$ according to Eq.\eqref{viscous-criteria-o},
which is much smaller than the adopted constant fluid time-step size $\Delta t^F = 2.0 \times 10^{-5}s$ demonstrating the efficiency of the present method.

\begin{table}
	\scriptsize
	\centering
	\caption{CPU time for viscous damping process in the simulation of hydrostatic water column on an elastic plate.}
	\begin{tabularx}{8.5cm}{@{\extracolsep{\fill}}lcc}
		\hline
		\quad & serial computation & parallel computation \\
		\hline
		CPU time (s) & $133.57$	& $26.89$  \\
		\hline
	\end{tabularx}
	\label{hydrostatic-time}
\end{table}
%
%
%
\section{Conclusions}
\label{conclusions}
In this paper, an efficient dynamic relaxation method for SPH method is proposed by imposing viscous damping to a dynamic system.
This method mainly includes the following aspects:
(a) a viscous damping term is added to the momentum conservation equation and two operator splitting schemes are introduced to discretize this term;
(b) a random-choice strategy is introduced to release the system from large damping; 
(c) a splitting CLL method is proposed to address the conflicts between different threads in shared-memory parallelization.
The implicit nature of the splitting scheme makes the discretization stable and large time-step size is allowable. 
The random-choice strategy helps to eliminate the path dependency, improve the convergence rate and also reduce much computational cost.
A various of test cases demonstrate the improved efficiency and performance of the 
present method for achieving the final converged state of a system compared to conventional SPH method.
Besides,
the proposed splitting CLL is thread-conflict free and makes it possible for implicit methods adopting share-memory parallel techniques to further improve computational performance.
Moreover, 
other implicit methods, 
e.g. the splitting scheme \cite{litvinov2010splitting} for highly dissipative problem and 
compact finite different scheme \cite{lele1992compact} for hyperbolic problems,
encounter the same problem of thread-conflict in shared-memory parallelization.
The present splitting CLL scheme also provides a practicable solution to these problems
by dividing computational domain into several blocks, which is the subject of our ongoing work.

\section*{Acknowledgements}
\addcontentsline{toc}{section}{Acknowledgement}
The first author is partially supported by Xidian University (China) and
the project of National Natural Science Foundation of China (NSFC) (Grant No:91952110).
C. Zhang and X.Y. Hu would like to express their gratitude to Deutsche Forschungsgemeinschaft for their sponsorship 
of this research under grant number DFG HU1572/10-1 and DFG HU1527/12-1. 
%
%

\bibliographystyle{elsarticle-num}
\bibliography{viscous-damping}
%
%
\end{document}